\documentclass[twocolumn]{aastex701}

\usepackage{graphicx}

\usepackage{amsmath}
\hypersetup{urlcolor=blue}

\newcommand\be{\begin{equation}}
\newcommand\ee{\end{equation}}

\begin{document}

\title{Correlations with Magnetic Activity in the Solar Near-Surface Shear Layer. II. Radial Shear}

\author[orcid=0000-0003-0172-3713, sname=Rabello Soares, gname=M. Cristina]{M. Cristina Rabello Soares}
\affiliation{W. W. Hansen Experimental Physics Laboratory, Stanford University, Stanford, CA, 94305-4085, USA}
\email[show]{csoares@sun.stanford.edu}
\correspondingauthor{M. Cristina Rabello Soares}

\author[orcid=0000-0002-6163-3472, sname=Basu, gname=Sarbani]{Sarbani Basu}
\affiliation{Department of Astronomy, Yale University, PO Box 208101, New Haven, CT 06520-8101, USA}
\email[]{sarbani.basu@yale.edu}

\author[orcid=0000-0002-0910-459X, sname=Bogart, gname=Richard]{Richard S. Bogart}
\affiliation{W. W. Hansen Experimental Physics Laboratory, Stanford University, Stanford, CA, 94305-4085, USA}
\email[]{rick@sun.stanford.edu}

\author[orcid=0000-0002-2307-0808, sname=Baldner, gname=Charles]{Charles S. Baldner}
\affiliation{W. W. Hansen Experimental Physics Laboratory, Stanford University, Stanford, CA, 94305-4085, USA}
\email[]{baldner@sun.stanford.edu}

\begin{abstract}
Using Helioseismic and Magnetic Imager ring-diagram measurements and building on the rotation-rate $\Omega$ inferences presented in Paper~I for depths of 1--17~Mm, we examine the properties of the dimensionless radial shear $\partial\ln\Omega/\partial\ln r$.
In the radial range overlapping global-mode analyses, the inferred shear agrees with previous results.
The near-surface shear layer exhibits a three-region shear structure with an enhanced-shear middle layer, and the largest residual variations occur in the two shallowest regions not accessible to global-mode analyses.
We parameterize the enhanced-shear layer by the depth of maximum shear, its amplitude, and its width, and find all three to be strongly correlated with a magnetic activity index; increasing activity corresponds to a shallower, stronger, and modestly narrower layer, indicating that the flows and magnetic fields are interconnected in these layers.
This behavior is consistent with expectations that sufficiently strong toroidal fields can enhance the near-surface rotational shear and with inferences of a near-surface toroidal-field concentration near the radius where we observe the strongest shear.
Moreover, the observed strengthening and upward shift of the strong-shear layer toward solar-cycle maximum suggest a corresponding solar-cycle dependence in the location of the strong toroidal field.
We also highlight the importance of finite-resolution effects and instrumental calibration in interpreting small variations in the results.
\end{abstract}


\section{Introduction} \label{sec:intro}

Shear layers in the Sun are thought to play a central role in generating and organizing magnetic fields through dynamo action. The deepest shear layer, the tachocline at the radiative–convective interface, remains a leading candidate for the seat of the solar dynamo \citep[e.g.][]{gilman2000}. However, there is growing evidence that shallower shear, concentrated above $\approx 0.95\,R_\odot$, may also contribute, or that multiple shear layers operate together \citep[e.g.,][]{brandenburg2005, pipin2011, vasil2024}. Characterization of the near-surface shear layer (NSSL) has therefore attracted considerable interest, but its structure and temporal variability are not yet fully understood, and its correlations with solar magnetic activity have not yet been fully explored.
Global helioseismology, based on spherical-harmonic decomposition, provides robust constraints at depth but loses sensitivity close to the photosphere; local helioseismology, by contrast, can probe depths as shallow as less than 1~Mm below the photosphere and allows independent analyses of the northern and southern hemispheres and of localized regions.

Assuming angular‑momentum conservation for an isolated fluid parcel undergoing radial motion gives $\Omega\propto r^{-2}$ and hence \citet{Foukal1975} postulated that the shear in NSSL, $\partial\ln\Omega/\partial\ln r$, should be $-2$. Helioseismic inversions, however, typically infer a less steep radial gradient, with values $\gtrsim -1$ at low and mid latitudes. That shallower gradient implies an increase in angular momentum throughout the NSSL toward the surface of $\gtrsim 5\%$, and therefore ongoing exchange of angular momentum.
The observations show that the shear varies with latitude: some studies report a modest weakening of the radial gradient at high latitudes, while others find the gradient becomes very small or disappears there \citep[e.g.,][]{barekat2014, antia2022}. Small temporal variations have also been observed, with some correlation to magnetic activity at the solar surface. Sunspots at the surface are confined to regions where the radial gradient is lower than average, around $0.99\,R_{\odot}$ 
\citep[][etc. ]{barekat2016, antia2022}. However, the pattern is reversed at $\approx 0.95\,R_\odot$ \citep[e.g.][]{antia2022}. The change with depth does not follow a clear trend and seems to vary in a non-uniform manner. The observations point to a subtle coupling among shear, flows, and magnetic fields that remains to be fully understood.

\begin{figure*}
\centering
\includegraphics[width=1.0\linewidth]{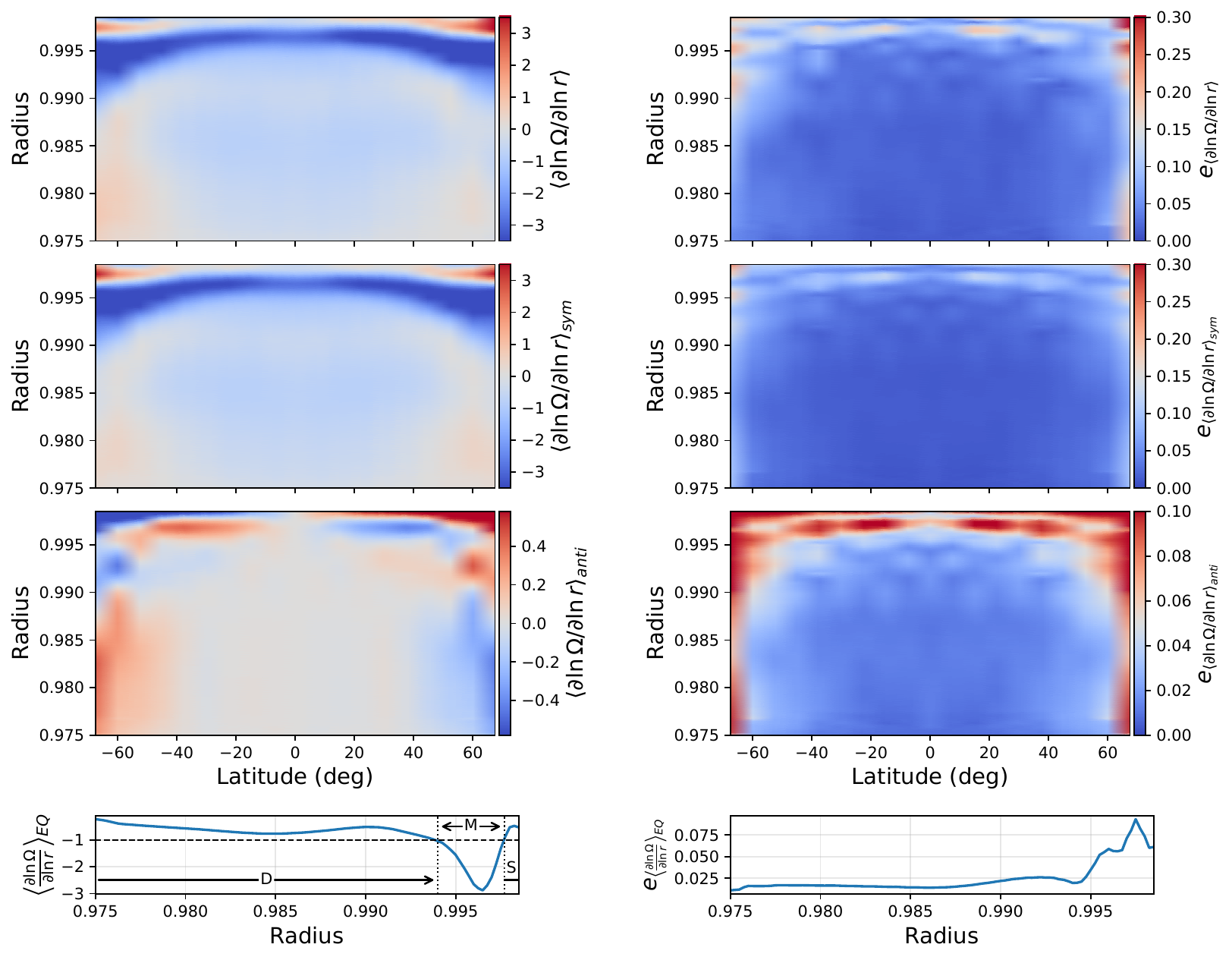}
 \caption{
 Columns on the left show the time-averaged logarithmic radial shear at the central meridian as a function of latitude and radius, while those on the right show the corresponding standard error on the mean (SEM) for the shear.  Top panels: the mean shear (left) and the corresponding SEM (right).  Second from top:  the north-south symmetric component of the shear (left) and the corresponding SEM (right).  Third from top: the antisymmetric component of the shear (left) and its SEM (right). Note that the antisymmetric component has a much smaller amplitude than the symmetric component, and hence we have plotted them on a different color scale.  Lowermost panels: the equatorial radial profile of the mean shear (left) and its SEM (right). The horizontal dotted line marks $\partial\ln\Omega/\partial\ln r = -1$. We mark layers D, M and S as defined by \citet{rabello2024}. Results were obtained with RLS inversions of 15-degree  ring diagrams; OLA inversions give similar results.
 }
    \label{fig:avg_shear}
\end{figure*}

By averaging over 12 years of data, \citet{rabello2024} had found that the NSSL can be divided into three fairly distinct regions (see Fig.~\ref{fig:avg_shear} lowermost left panel): a deeper, larger region with a small shear (layer D), steepening toward the surface; a narrow middle layer, layer~M,  with a strong shear that first increases and then decrease ---  the maximum gradient is approximately three times larger than in layer D; and a layer very close to the surface (layer~S), where the logarithmic gradient is close to zero around 1.6~Mm depth ($0.9977\,R_{\odot}$) but becomes steeper again toward the surface (layer S). At layer D, we estimated $\partial\ln\Omega/\partial\ln r=-0.5$ at $0.97\,R_{\odot}$ and $-0.8$ at $0.99\,R_{\odot}$, in general agreement with previous work. The latitudinal variation changes with depth in layer D: the gradient goes from $-0.5$ at the equator to about zero or slightly positive at $60^\circ$ latitude. Near the top of layer D, i.e. at approximately $0.99\,R_\odot$, close to layer M, the gradient goes from $-0.8$ at the equator to $-3$ at high latitudes, i.e., the opposite behavior compared to some previous works
\citep{corbard2002, zaatri2009, barekat2014, barekat2016, reiter2020}. The pronounced steepening in the middle layer, i.e., layer~M, had been hinted at in earlier studies.  \citep{basu_etal1999, zaatri2009, komm2022}, however, \citet{rabello2024} were able to better localize the region and found that the maximum of the absolute value of the shear to be at a depth of at 2.4~Mm ($0.9965\,R_{\odot}$), where $\partial\ln\Omega/\partial\ln r=-3$. However, they showed that the poor resolution in these layers implied that  the maximum could be located closer to the surface, around 1.5~Mm depth. Inside layer M, the gradient goes from a steep value at the equator to a still negative but much smaller value, around $-1$, at higher latitudes.

This is the second of two papers about the characteristics of the NSSL. In Paper~I (Rabello Soares et al., submitted) we examine the rotation rate and its variation with time. We found that the torsional oscillation signal can be found even in the shallowest layers that we could resolve, a depth of about 1~Mm. We also found that the cumulative zonal displacement inferred from the residual flows is significantly correlated with magnetic activity at some latitudes, with multi-year temporal offsets.
In this work we pursue three complementary goals: (1) to characterize the time‑averaged, dimensionless, radial shear of the rotation rate and its variation with latitude; (2) to quantify the temporal variability of the shear; and (3) to determine how the radial shear couples to magnetic activity. Section~\ref{sec:data} describes the data and methods. Section~3 presents our results and is structured as follows: Section~\ref{subsec:avshear} presents the time‑averaged structure, Section~\ref{subsec:timeshear} analyzes temporal changes, and Section~\ref{subsec:mag} quantifies correlations with magnetic activity. The section on correlations with activity is further organized into (i) characterization of the NSSL shear, (ii) activity‑dependent changes, and (iii) a quantitative correlation analysis of layer‑M properties (peak amplitude, location, width) analyzed as functions of surface magnetic field. Finally, Section~\ref{sec:conclusions} summarizes the main findings and discusses their physical implications.

\section{Data and Analysis Technique} \label{sec:data}

We use Dopplergrams from the Helioseismic and Magnetic Imager\citep[HMI:][]{hmi}, an instrument on board the Solar Dynamics Observatory (SDO), and apply the same analysis procedures as in Paper~I. The data span the time interval 2010 May 1 (Carrington rotation 2096) through 2025 Feb 3 (Carrington rotation 2293).
For this work we use tiles that have a diameter of  $15$ degrees and $30$ degrees in heliographic coordinates. The centers of  $15$-degree tiles were separated by $7.5$ degrees, while the $30$-degree tiles were separated by $15$ degrees. The $15$-degree tiles were tracked for about 8.9 hours, $30$-degree tiles for 57.6 hours. In this work we only use tiles on the central meridian and along the equator.  

Next, we determined the   spatial-temporal power spectra of the resultant data cubes. The power spectra were then averaged over a Carrington rotation --- i.e., we averaged 24 spectra of the $15$-degree tiles, and 12 of the $30$-degree tiles.
The rotation-averaged spectra were then fitted using module \texttt{rdfitc} \citep{Bogart_pipeline} that fits the model of \citet{basu_antia1999} to obtain the shift \(u_x(\theta,t,\ell,n,\nu)\) caused by the zonal (east-west) velocities for each detected mode of degree $\ell$, radial order $n$ and temporal frequency $\nu$. The $u_x$ parameter includes the rotational velocity (modulo the tracking rate) and any other prograde or retrograde zonal flows. 
To suppress short‑timescale variations observed by \citet{Bogart2023} while preserving solar‑cycle trends, the fitted $u_x$ time series for each $(\ell,n)$ pair at a given latitude $\theta$ were smoothed with a 1‑yr running mean (13 Carrington rotations), advanced by one rotation between successive averages. 

A depth profile of the zonal velocity $U_x(\theta,t,r)$ for each averaged power spectrum was obtained by inverting the flow parameter $u_x$. The inversions were carried out using two independent inversion techniques: Optimally Localized Averages (OLA) \citep[module \texttt{rdvinv};][]{Bogart_pipeline}, 
and the Regularized Least Squares (RLS) using the implementation of \citet{basu_etal1999}.
For each method, we determined the regularization/trade-off parameter for the inversions by balancing noise amplification against localization \citep{rs1999}. 
We use both OLA and RLS inversions as an internal consistency check; because the methods are complementary \citep[e.g.,][]{sekii1997}, their agreement increases confidence in the inferred results. The inversion results $U_x$ can be converted to the rotation rate $\Omega$ for each tile and each depth easily using $\Omega=U_x / (r\cos\theta)$ and adding the tracking rate.

Since the NSSL is generally characterized by the logarithmic radial derivative of the rotation rate $\partial\ln\Omega/\partial\ln r$, the inverted values of the zonal velocity $U_x$ were  used to determine the radial shear $\partial U_x/\partial r$, and then,  using the known tracking rate, converted to the logarithmic derivative of the rotation rate $\Omega$ at each latitude and radius, as was done by \citet{rabello2024}. Henceforth, we will refer to this logarithmic radial derivative as the ``shear''.

The motivation for fitting rotation-averaged spectra, rather than simply downloading and using the standard $U_x(\theta,t,r)$ product provided by the HMI pipeline, is that averaging the spectra before fitting significantly reduces realization noise, and decreases uncertainty, and it also allows the smaller peaks to rise from the noise. This improves the inversions by reducing oscillatory artifacts in the inferred depth profiles driven by error correlations in the fitted mode parameters \citep{howe_thompson}. Such spurious oscillations hinder interpretation since they are amplified when computing the radial rotation gradient.

We quantify the magnetic field  strength as a magnetic activity index \citep[MAI;][]{Bogart_pipeline}. The MAI for each tile  was determined from HMI magnetograms for the same tracked regions used in the ring-diagram analysis by integrating the unsigned magnetic flux above a 50~G threshold \citep{Bogart_pipeline}; the threshold was chosen based on the noise characteristics of HMI magnetograms. As with the fitted velocity parameters, the MAI values were averaged within each Carrington rotation and smoothed with a 1-yr running mean, shifted by one rotation, in the same manner as the $U_x$ time series.
This enables a robust correlation of our helioseismic results with the surface magnetic field strength.

\section{Results}
\label{sec:results}

\subsection{Average shear in the NSSL}
\label{subsec:avshear}

\begin{figure*}
\includegraphics[width=\linewidth]{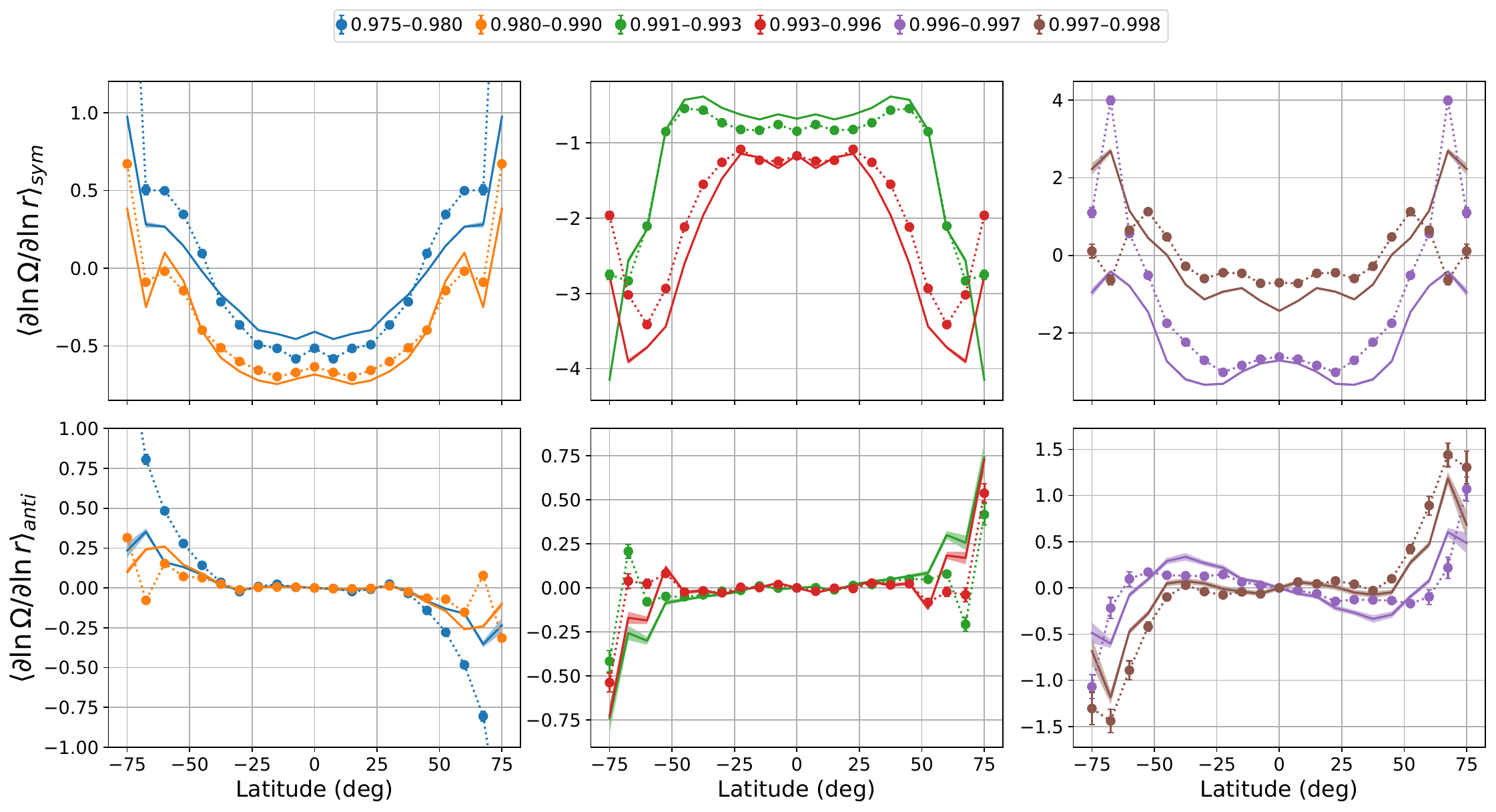}
\caption{The north-south symmetric, i.e., equatorially symmetric, component of the shear plotted as a function of latitude averaged over different radial extents. The colors indicate the radius interval over which the average was taken, and these are indicated in the legend.  The lower panels show the corresponding north-south antisymmetric component of the shear. OLA results are shown as dotted lines with filled circles, and RLS results as full lines. The error bars on the circles and the shaded bands around the curves indicate the standard error of the mean. Fig.~\ref{fig:avg_shear} shows the full depth–latitude maps of the radial shear.
    }
    \label{fig:mean_rotation_shear_sym}
\end{figure*}

We characterize the NSSL by the logarithmic radial shear, $\partial\ln\Omega/\partial\ln r$, computed from the inverted $U_x$ as described in Section~\ref{sec:data}. 
Before we delve into the temporal variations of the shear, we show  the time-averaged shear in Fig.~\ref{fig:avg_shear} (left column).  As can be seen, the most prominent feature is a strong shear (in blue). This feature is narrow at the equator; the band  strengthens and moves slightly deeper, and widens at higher latitudes. This behavior is robust across inversion methods. In deeper layers, the shear has a value close to $-1$; this agrees with global-mode results \citep[see e.g.,][etc.]{barekat2014, antia2022}.  There is a small north-south asymmetry in the results, which is about six times smaller than  the symmetric component.  The standard error of the mean, shown in the right column of Fig.~\ref{fig:avg_shear}, is quite small.

To explore the latitudinal dependence of the shear, in  Fig.~\ref{fig:mean_rotation_shear_sym} we show  the time-averaged radial shear, $\langle\partial\ln\Omega/\partial\ln r\rangle$ plotted  as a function of latitude, averaged over selected radial intervals. 
For $r\lesssim 0.990\,R_\odot$, $\langle\partial\ln\Omega/\partial\ln r\rangle_{\mathrm{sym}}$ is negative ($\sim -0.6$) at the equator and approaches zero at latitudes of $45^\circ$--$60^\circ$, indicating nearly depth-independent rotation there; this is broadly consistent with previous studies 
\citep[][etc.]{corbard2002, zaatri2009, reiter2020}.
Other studies also report that the magnitude of the shear decreases toward high latitudes, but only modestly, by $\sim20$\% \citep{barekat2014, barekat2016, antia2022, mandal2025}. There is no consensus on the results at even higher latitudes, and our OLA and RLS inferences also disagree there. \citet{mandal2025} report that the shear approaches zero near $75^\circ$ latitude; however, \citet{antia2022} found a steepening of the shear at latitudes greater than $60^\circ$, in qualitative agreement with our results.

The symmetric component of the average shear over the range $0.991$--$0.996\;R_\odot$ shows a transition in its latitudinal dependence relative to deeper layers, with two distinct trends: (i) at the equator the shear becomes more negative with depth ($\sim-0.6$ to $\sim-2$); and (ii) at fixed depth it decreases (becomes more negative) by $\sim 2$ from the equator to high latitudes. 
Thus, in this mid range the shear magnitude increases toward high latitudes, opposite to the deeper trend. At shallower depths, 
 $r\in[0.996,0.998]\;R_\odot$, the latitudinal dependence reverses: the shear is negative at the equator and increases by $\sim 2$ between the equator and $60^\circ$ (i.e., becomes more positive), reaching near-zero or positive values by $60^\circ$.
The changes in the latitudinal dependence as a function of radius are consistent with those found by \citet{rabello2024}; the last two intervals listed above approximately sample layers M and S, and the trends are the same as those  reported in our previous work (Fig.~6 in \citealt{rabello2024}). 
The latitude dependence clearly changes with depth, likely explaining discrepancies among previous results that sample different depth ranges.

The antisymmetric component, $\langle \partial\ln\Omega/\partial\ln r\rangle_{\mathrm{anti}}$, increases in magnitude with absolute latitude, $|\theta|$, 
 and has its largest amplitude close to the surface ($r \gtrsim 0.996\,R_\odot$), for both OLA and RLS.
\citet{rabello2024} compared this equatorially antisymmetric ($(N-S)/2$) shear with the east–west antisymmetric component relative to the central meridian ($(E-W)/2$) and found markedly different behavior, suggesting that the observed north-south asymmetry is not dominated by the so-called center-to-limb systematics \citep[see][]{zhao2012}.

\subsection{Time variation of the shear}
\label{subsec:timeshear}

\begin{figure}
    \includegraphics[width=\columnwidth]{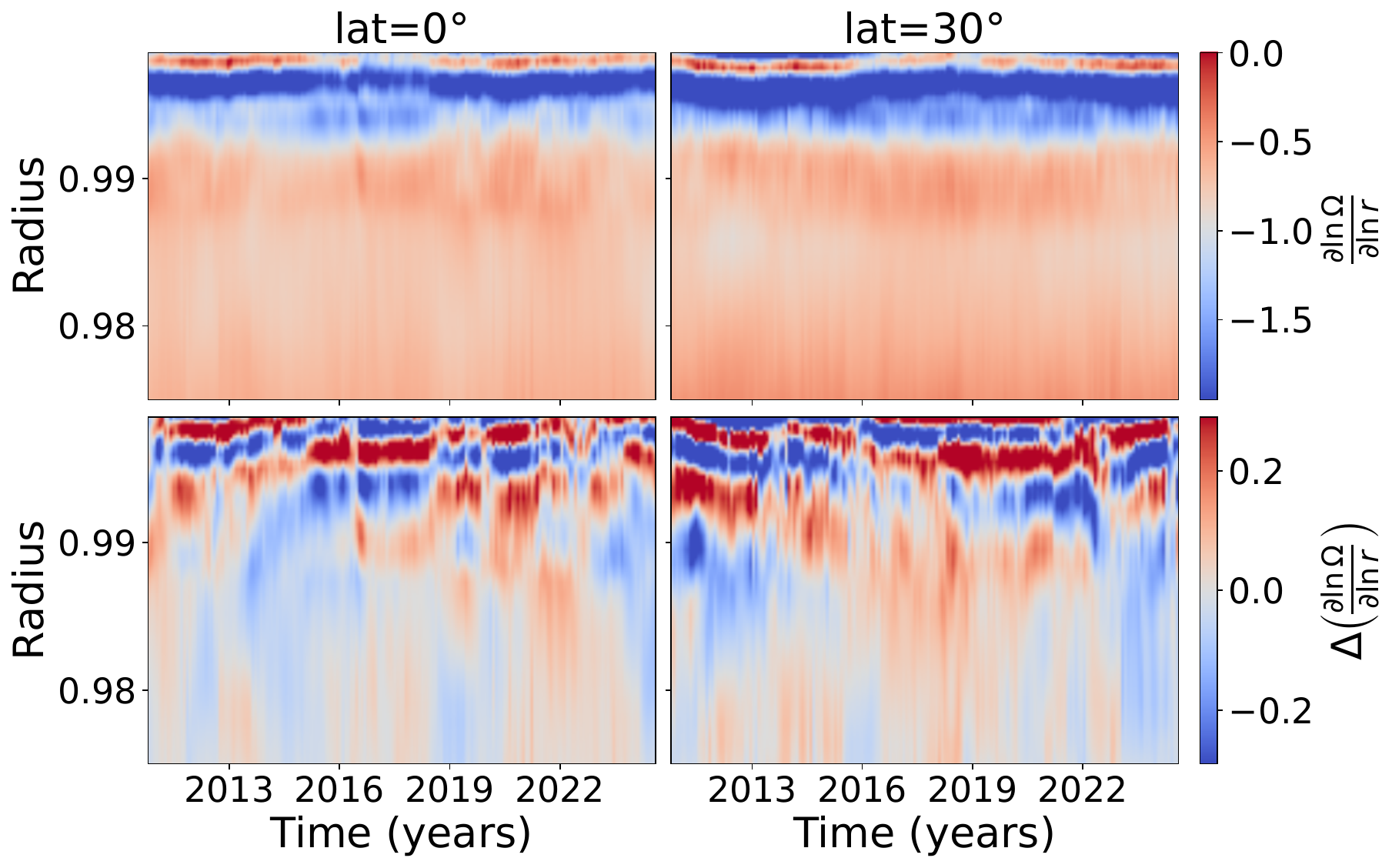}
    \caption{Top: The north-south symmetric component of the shear plotted as a function of depth and time for two different latitudes. These panels share a common color scale. Bottom: The symmetric component of the residuals in shear after removing the time-averaged shear at each latitude and depth from the top panel;  these panels share a common color scale. Both inversion methods and tile sizes yield consistent results within their mutual sensitivity ranges, although we have only shown RLS results.
    }
    \label{fig:shear_depth_time_sym}
\end{figure}

\begin{figure}
        \includegraphics[width=\columnwidth]
    {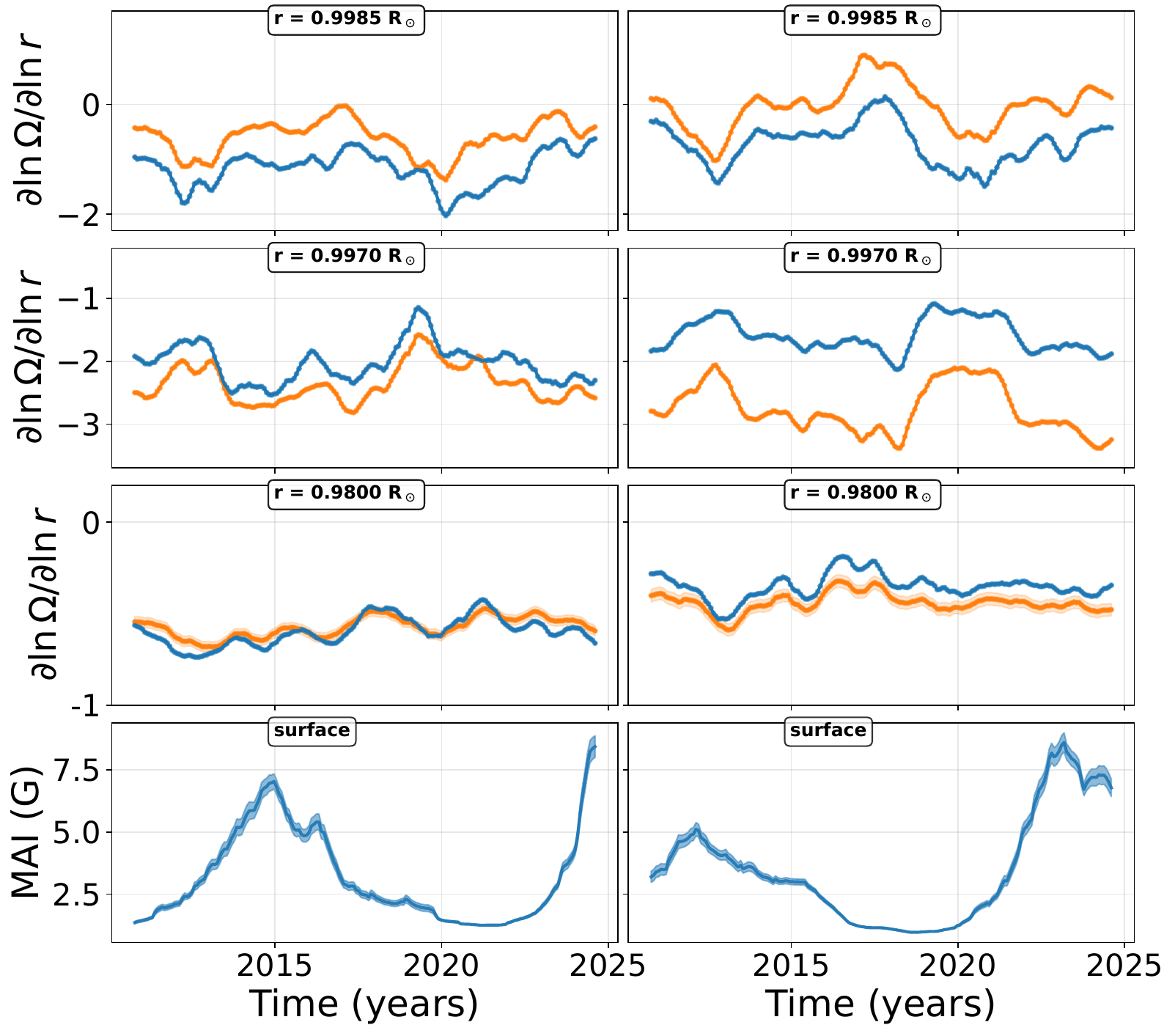}
    \caption{The north-south symmetric component of the shear obtained from 15° tiles,  using OLA (blue) and RLS (orange) inversions, shown as a function of time at two latitudes (equator, left column; $30^\circ$, right column) and three radii: $r=0.9985\,R_\odot$ ($\approx 1$~Mm below the photosphere, i.e., layer S; top), $r=0.9970\,R_\odot$ ($\approx 2$~Mm, layer M; middle), and $r=0.9800\,R_\odot$ ($\approx 14$~Mm, layer D; bottom). The solid curves show the gradients smoothed with a 7-rotation running mean after removing points that were more than $3\sigma$ from the average.  Shaded bands show the uncertainty and are plotted only for the RLS inversions; they are visible only at the deepest level. The lowermost panels show the symmetric component of the MAI at the corresponding latitude.
    }
    \label{fig:3layers}
\end{figure}

\begin{figure}
    \includegraphics[width=\columnwidth]{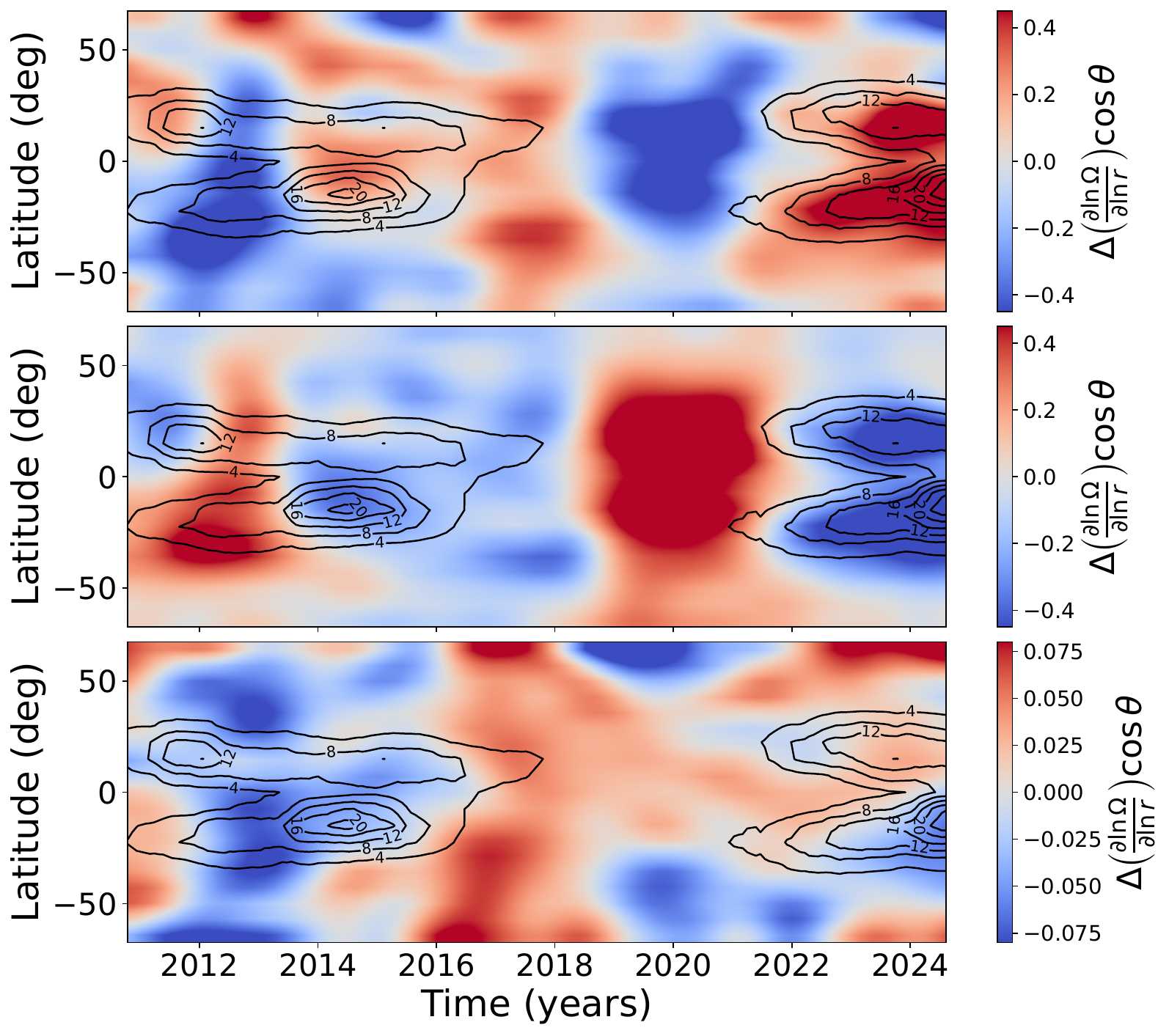}
    \caption{
    Residuals in the shear for the results obtained with $15$-degree tiles obtained using RLS inversion, plotted as functions of time and latitude at $r = 0.9985\,R_\odot$ ($\sim 1$~Mm below the photosphere; top), 
     $r = 0.9970\,R_\odot$ ($\sim 2$~Mm deep; middle), and $r = 0.9800\,R_\odot$ ($\sim 14$~Mm deep; bottom). The residuals were smoothed with a two-dimensional Gaussian filter ($\sigma=7.5^\circ$ in latitude and 6 Carrington rotations in time) to obtain a smooth figure. Contours of the photospheric magnetic activity index (MAI; 4–24 G) are overplotted. Similar results are obtained with OLA inversions. Note that the bottom panel uses a different color scale; near-surface residuals are substantially larger (see Fig.~\ref{fig:shear_depth_time_sym}).
    }
    \label{fig:res_lat_time}
    \end{figure}
\begin{figure}
	\includegraphics[width=0.9\columnwidth]{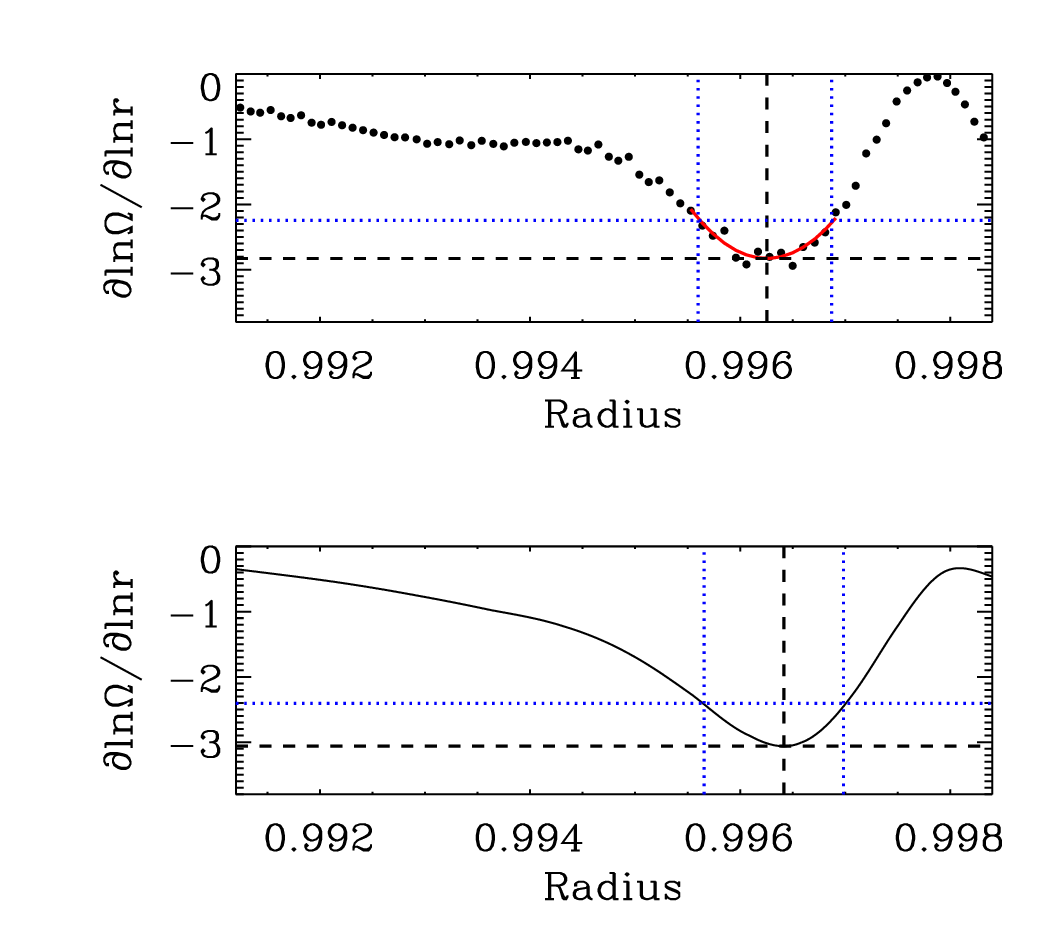}
    \caption{
    Determination of the maximum rotational shear and corresponding width (\(\mathrm{FW80}\)) using OLA (top) and RLS (bottom) for the 15-degree tile at the equator for the 1-yr set centered at CR 2113.   The vertical dashed black lines mark $r_{\max}$ and the horizontal dashed black lines mark $A_{\max}$. The horizontal dotted blue line marks 0.8~$A_{\max}$ and the vertical dotted blue lines indicate the radii whose separation defines $\mathrm{FW80}$.
    }
    \label{fig:rmax_amax_definition}
\end{figure}

We study the time variation in the shear in a manner similar to that of the rotation rate itself.  In Fig.~\ref{fig:shear_depth_time_sym} we show the north-south symmetric component of the shear as a function of time at two latitudes. The shear varies in time; this was also seen in global-mode analysis \citep{antia2022}. 
However, we find that the most significant changes occur at depths that are inaccessible to the global-mode data sets that are normally available. The lower panels of the figure show the residuals left after the mean shear is subtracted. This confirms that the largest changes are in the shallowest layers, and we also find that the changes in the layers that show the largest shear (layer M of \citealt{rabello2024}; see Fig.~\ref{fig:avg_shear}) generally have the opposite sign than the changes in the shallowest region (layer S). The changes in the deeper part of the NSSL (layer D) are much less pronounced. In order to get a clearer view of the time variation, we have plotted the shear at three different depths ($\sim 1$, $\sim 2$, $\sim 14$~Mm)  as a function of time at the equator and a latitude of $30^\circ$; we also show the variation of the MAI at these two latitudes (Fig.~\ref{fig:3layers}). Note that the changes in the shear are in opposite directions in layers S (upper panel) and M (second from top). 

Fig.~\ref{fig:res_lat_time} provides a complementary view and shows the residuals of the shear plotted as functions of time and latitude at three depths, $\sim 1$~Mm (top), $\sim 2$~Mm (middle) and $\sim 14$~Mm (bottom),  with the corresponding photospheric magnetic activity index (MAI) overplotted for comparison. Note that the residuals in the two top panels, which show very shallow layers, are larger than those in the bottom panel. At $0.997\,R_{\odot}$ (middle panel) our results are consistent with previous work at similar radii ($0.99\,R_\odot$), which found that sunspots tend to lie in regions where the gradient is below average \citep{barekat2016, antia2022, komm2022}. At the shallowest depth shown (upper panel), the residuals show the opposite behavior ---  they are enhanced (i.e., have the opposite sign) in regions of strong magnetic activity.

\subsection{Correlation with magnetic activity}
\label{subsec:mag}

\subsubsection{Characterizing the NSSL shear}

The residuals of the NSSL shear vary with depth, latitude, and time, and their sign relative to the photospheric magnetic field (which we describe in terms of the MAI), and hence, the residual maps alone do not provide a clear picture of the changes in the NSSL and the factors affecting its properties. Since the middle layer of the NSSL shows the largest shear and also the largest changes, this is the layer that we concentrate on. For this study,  we quantify the evolution of the strong shear by tracking its central depth, peak amplitude, and radial width as functions of time for each latitude band, and we test these diagnostics for correlations with magnetic activity over the solar cycle. 

As mentioned earlier (and shown in the lowermost panel of Fig.~\ref{fig:avg_shear}) the NSSL can be divided into three regions, and that the central region has the strongest gradient. For each latitude band and Carrington rotation,  we locate the position where the radial shear has the largest amplitude in the radius range  $r = 0.995 \pm 0.002\,R_\odot$ (see Fig.~\ref{fig:rmax_amax_definition}).
For the OLA solutions, we fit a second-degree polynomial to the rotational gradient over the local neighborhood 
and define $r_{\max}$ as the radius where the derivative of the fitted polynomial vanishes; $A_{\max}$ is the value of the shear at $r_{\max}$.
We express this location, for convenience, as a depth, $d_{\max} \equiv (1-r_{\max})\,R_\odot$, and report $d_{\max}$ in Mm.
For the RLS solutions, represented by splines, we interpolate the RLS gradient over the same radius range  and locate the value with the largest absolute value
in the  profile to obtain $r_{\max}$, the location, and $A_{\max}$, the value of the gradient. To analyze the radial extent of the strong-shear layer, we define its width, $\mathrm{FW80}$, as the difference between the radii immediately to the right and left of $r_{\max}$ where the shear equals $0.8\,A_{\max}$, using linear interpolation.

\subsubsection{Variation with magnetic activity}

In addition to the strong latitudinal variation (see Fig.~\ref{fig:avg_shear}), $d_{\max}$, $A_{\max}$, and \(\mathrm{FW80}\) vary with time and are also correlated with the surface magnetic activity index (MAI). 

We find that if we remove the average latitudinal trend in $d_{\max}$, at low latitudes (between $\pm 22.5^\circ$), the Pearson and Spearman correlation coefficients between $d_{\max}$ and MAI are in the range $-0.5$ to $-0.8$ ($p<10^{-3}$), with the strongest correlation near $15^\circ$ latitude. The corresponding coefficients of determination, $R^2$, from linear fits range from $\simeq 0.3$ to $0.5$. The relationship weakens substantially at higher latitudes; for example, at $30^\circ$ latitude the Pearson correlation coefficient is $-0.2$ and $R^2\simeq 0.04$. Outside the sunspot belts ($>30^\circ$), the MAI is typically $\lesssim 3$ G and varies by $\lesssim 2$ G over the cycle, which likely limits sensitivity to any activity dependence. A similar behavior is found for $A_{\max}$, as well as $\mathrm{FW80}$: statistically meaningful correlations with MAI are confined to the activity belts, while at $\gtrsim 30^\circ$ the correlations and $R^2$ values are negligible.

\subsubsection{Quantifying the correlation with activity in the active latitudes}

To quantify the behavior within the activity belts, we performed a combined regression across low latitudes from $22.5^\circ$S to $22.5^\circ$N. For each estimated parameter, $d_{\max}$, $A_{\max}$, and \(\mathrm{FW80}\), we used results at seven latitudes ($-22.5^\circ$, $-15^\circ$, $-7.5^\circ$, $0^\circ$, $7.5^\circ$, $15^\circ$, $22.5^\circ$) and fitted a log-linear model, where we assume that the slope of the relation is independent of latitude, but that the intercept may have a latitudinal dependence. The specific model we fit is  
\begin{equation}
Y(\theta,t)=c_{0,Y}(\theta)+c_{1,Y}\,\ln\!\left[\mathrm{MAI}(\theta,t)\right],
\label{eq:lat_pooled}
\end{equation}
where $Y(\theta,t)$ denotes either $d_{\max}$, $A_{\max}$ or \(\mathrm{FW80}\) at a selected latitude and a given epoch $t$. The parameter $c_{0,Y}(\theta)$ is the latitude-dependent intercept and $c_{1,Y}$ is a slope common to all included latitudes. Note that we use MAI in units of Gauss to calculate $\ln(\mathrm{MAI})$ in Eq.~\ref{eq:lat_pooled}. The coefficients were estimated by the ordinary least squares method using 1,300 values of each parameter pooled across the seven latitudes. Fits were performed separately for results obtained with OLA and RLS inversions.

\begin{figure}
	\includegraphics[width=\columnwidth]{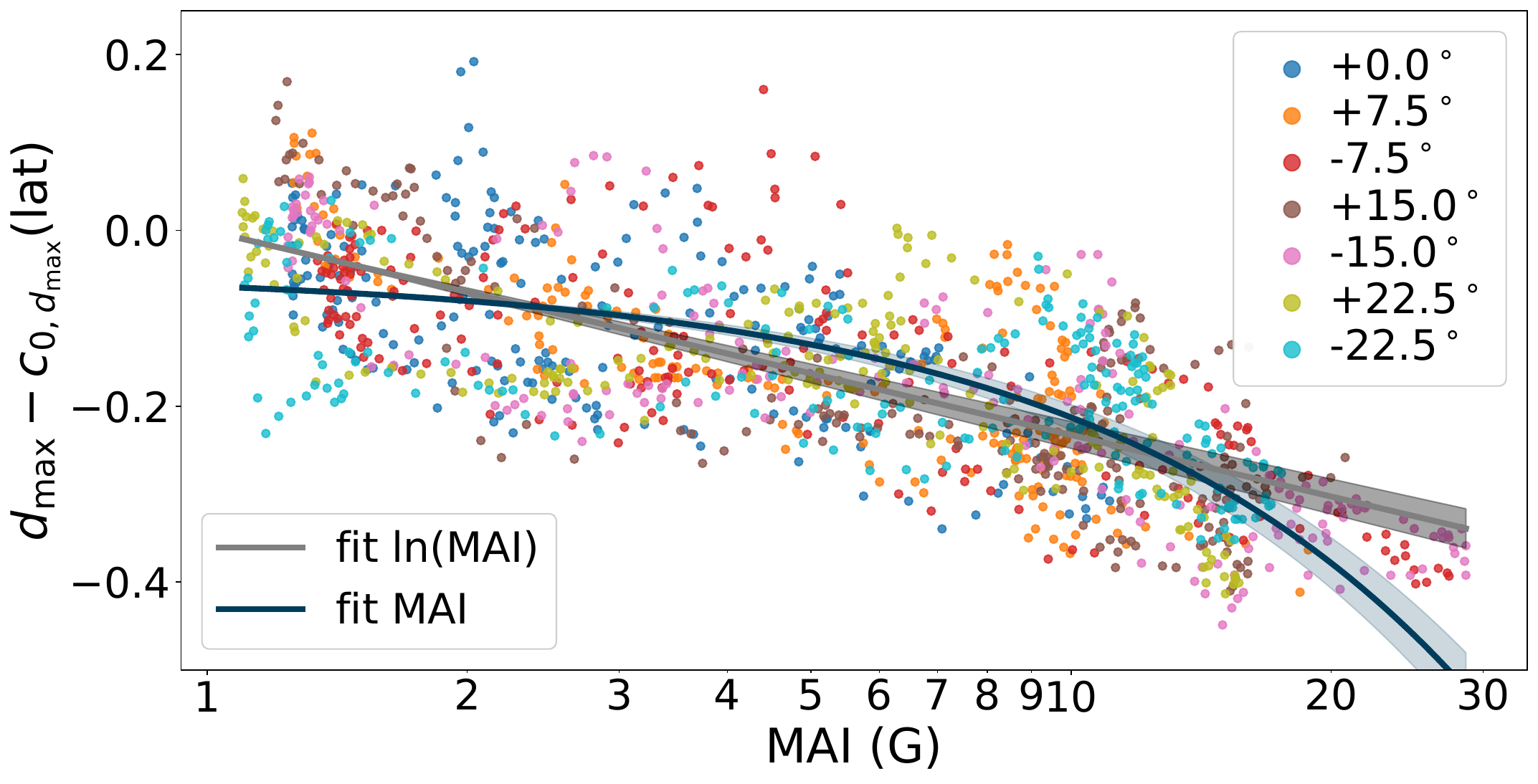}
    \includegraphics[width=\columnwidth]{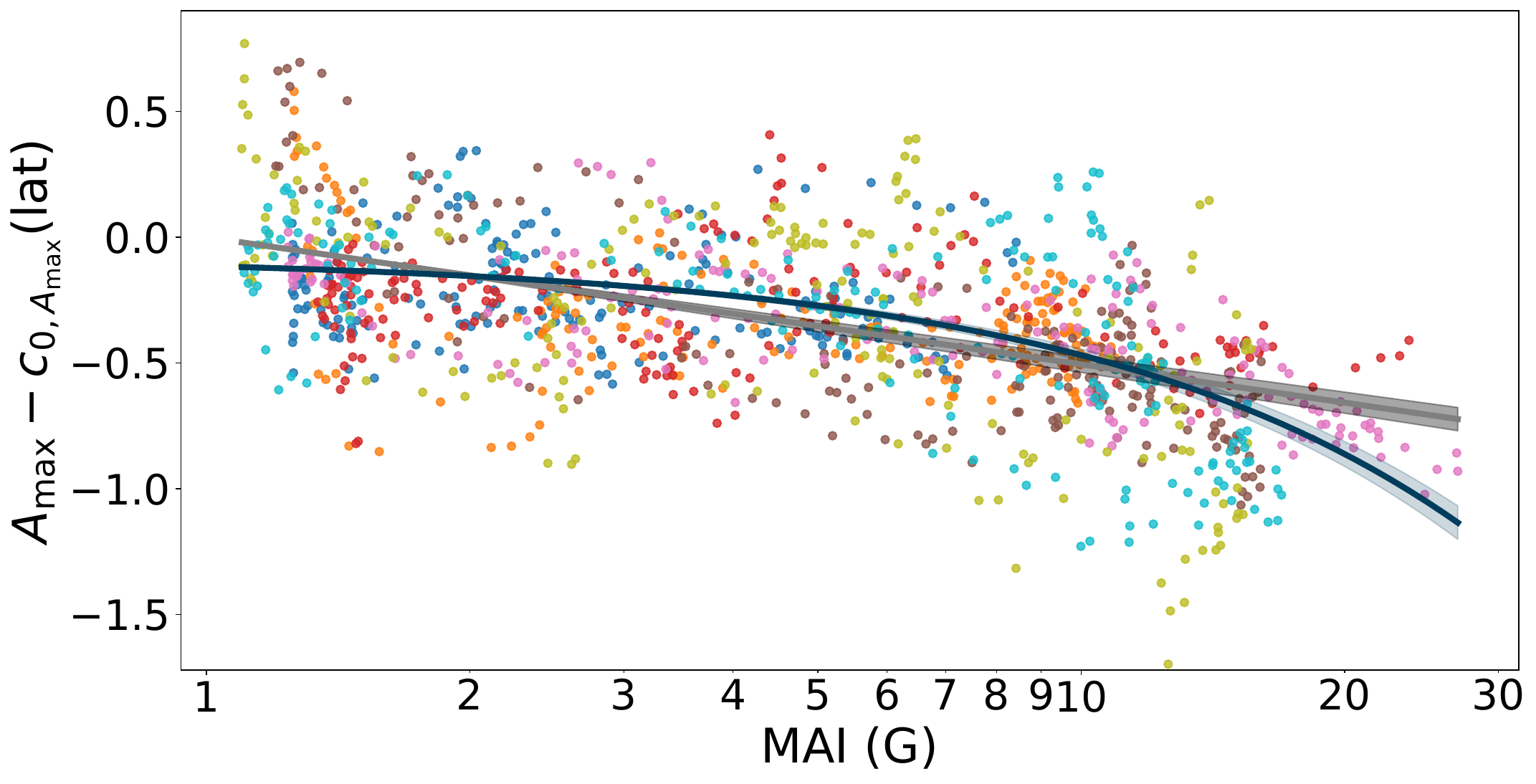}
    \includegraphics[width=\columnwidth]{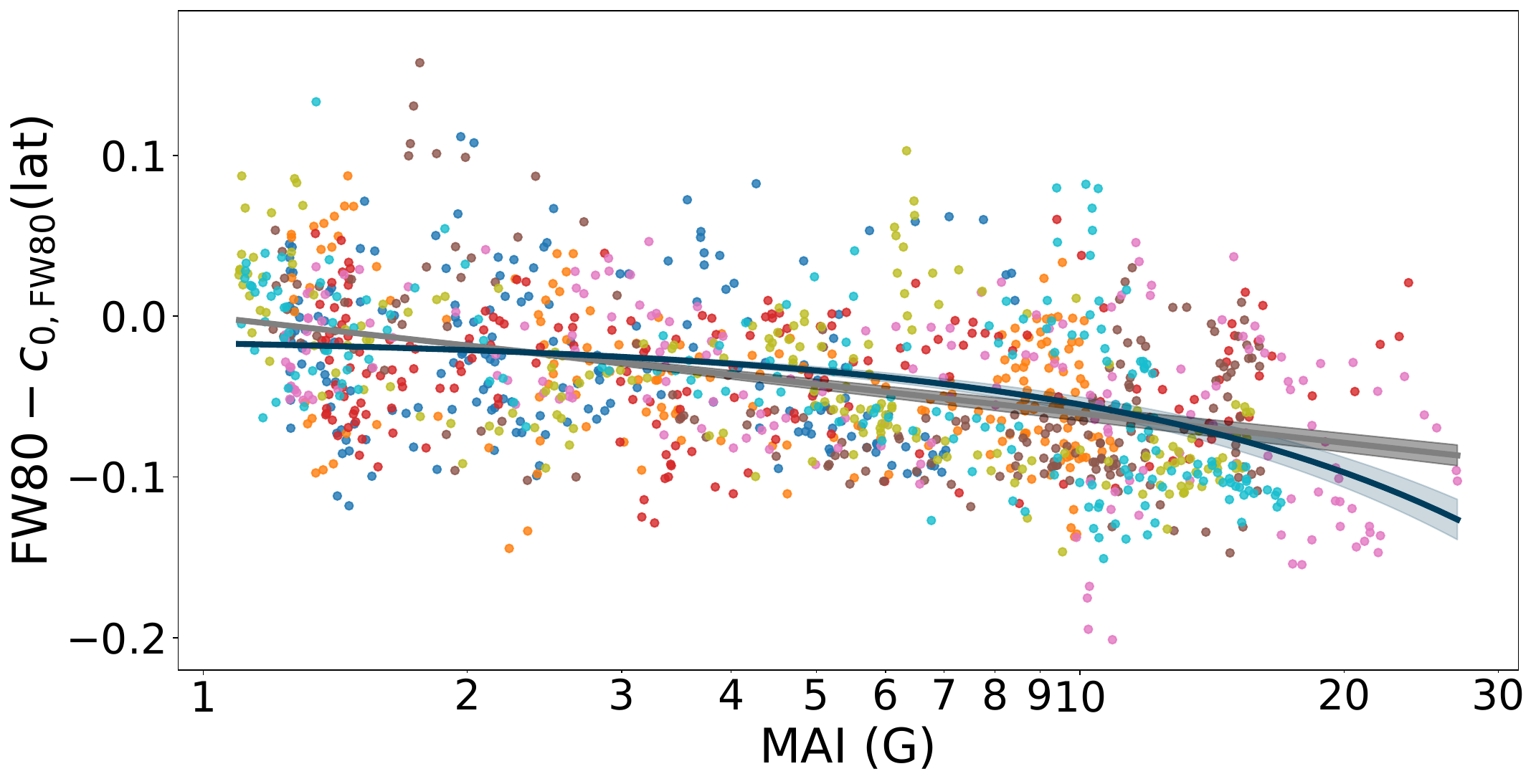}
    \caption{The quantities \(d_{\max}\), \(A_{\max}\), and \(\mathrm{FW80}\) obtained with RLS inversions and plotted as a function of MAI; OLA results are very similar. Note that the abscissa is on a logarithmic scale. Each quantity has been corrected for latitude offsets by subtracting the fitted intercepts \(c_{0,Y}(\theta)\). 
    The solid gray lines show the best-fit linear regression versus \(\ln(\mathrm{MAI})\), and dark blue lines show the corresponding fits versus MAI (and hence are curved in the figures).
    }    
    \label{fig:rmax_mai_clatns_mai}
\end{figure}

\begin{figure}
	\includegraphics[width=\columnwidth]{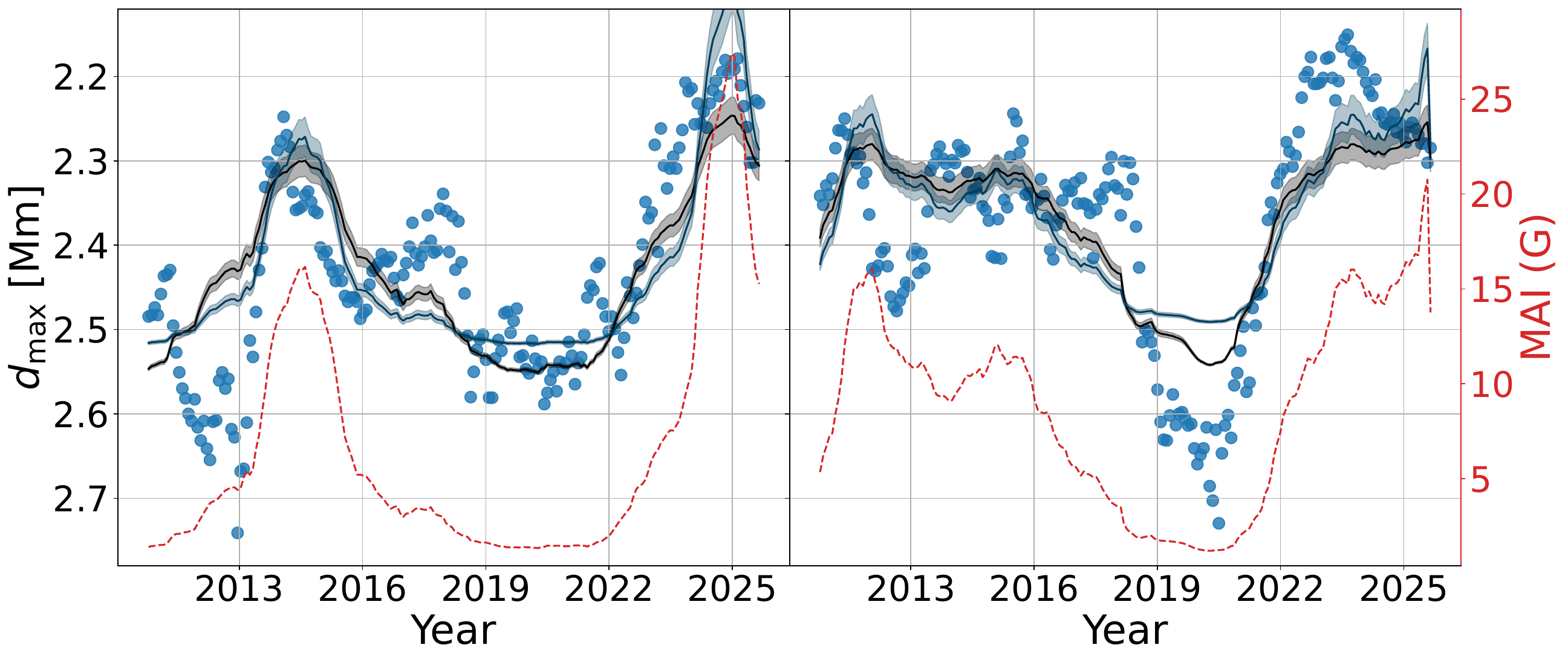}
    \includegraphics[width=\columnwidth]{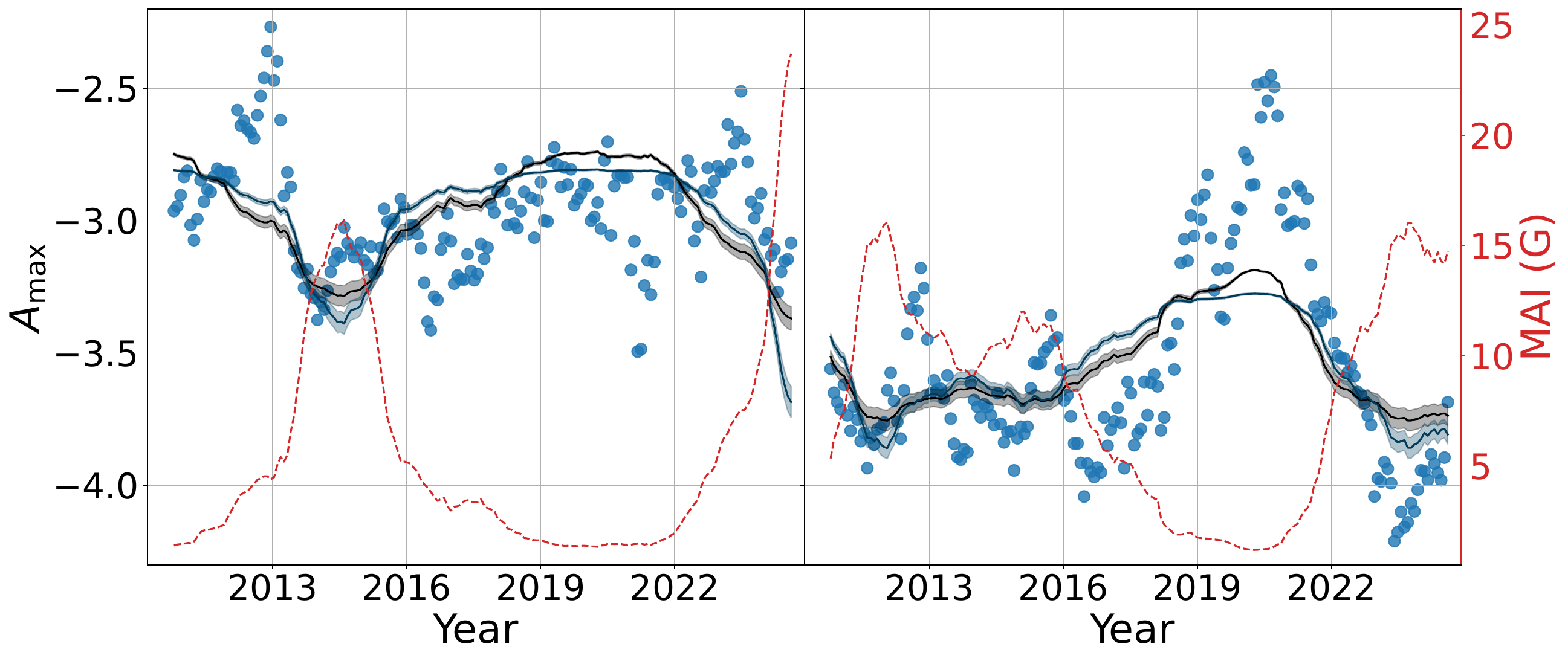}
    \includegraphics[width=\columnwidth]{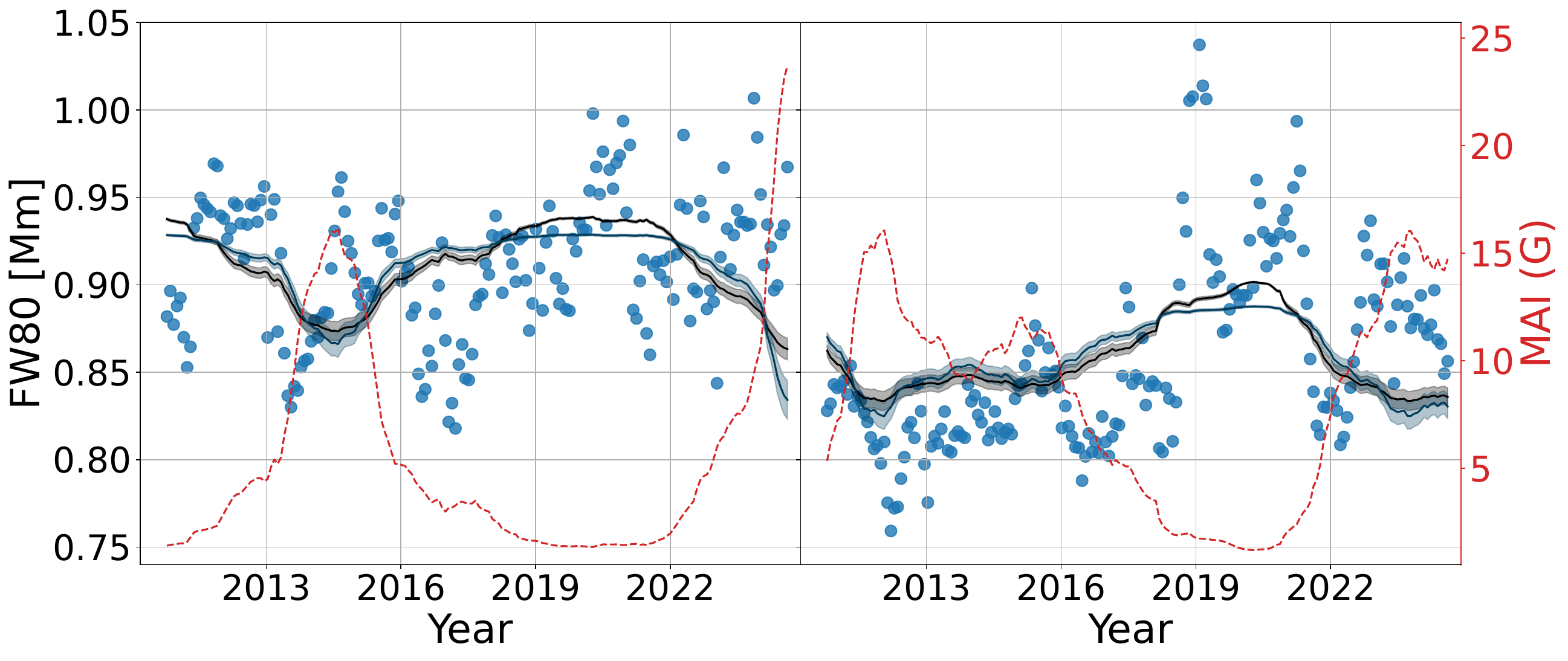}
    \caption{The parameters \(d_{\max}\), \(A_{\max}\), and \(\mathrm{FW80}\) plotted as a function of time for the RLS inversion; the OLA results are very similar. 
    The left column shows results at $7.5^\circ$S and the right column at $15^\circ$N. 
    The solid gray lines show the best-fit regression versus \(\ln(\mathrm{MAI})\), and the dark blue lines show the corresponding fits versus MAI. The red curve shows the MAI; the MAI values can be read from the right-hand axis.  
    }
    \label{fig:rmax_mai_clatns_time}
\end{figure}

Fig.~\ref{fig:rmax_mai_clatns_mai} shows $d_{\max}$, $A_{\max}$, and \(\mathrm{FW80}\), after subtracting the corresponding latitude-dependent intercept $c_{0,Y}(\theta)$, plotted as a function of MAI, together with the fitted model. In Fig.~\ref{fig:rmax_mai_clatns_time}, we show the time-variation of these quantities at two latitudes, $7.5^\circ$S and $15^\circ$N, again with the corresponding model curves overplotted.

\paragraph{\bf Behavior of $d_{\max}$}
 Fig.~\ref{fig:rmax_mai_clatns_mai} indicates that $d_{\max}$ moves toward shallower layers as MAI increases.
The latitude-specific intercepts $c_{0,d_{\max}}(\theta)$ are in the range 2.5--2.7~Mm across OLA and RLS (for $|\theta|\le 22.5^\circ$). For the model in Eq.~\ref{eq:lat_pooled}, $R^2 = 0.50$. The $t$ ratios (estimate divided by standard error) are $>160$ for $c_{0,d_{\max}}(\theta)$ and 13 for the common slope $c_{1,d_{\max}}$, indicating that both are strongly significant. 

To verify that the slope is not affected by $c_{0,d_{\max}}(\theta)$, we subtracted the fitted $c_{0,d_{\max}}(\theta)$ from $d_{\max}$ and refitted a slope-only regression; the slope was essentially unchanged, and this partial model yields $R^2 = 0.80$ (Table~\ref{tab:dmax_corr}), implying that $\ln(\mathrm{MAI})$ explains most of the variance once latitude offsets quantified by $c_{0,d_{\max}}(\theta)$ are removed. The relationship is also supported by the correlation measures (Pearson $r \simeq -0.7$, Spearman $\rho \simeq -0.7$; $p \ll 10^{-6}$ in both cases) and by the slope $t$-ratio ($\simeq -25$; $p \ll 10^{-6}$).

Diagnostic tests on the full model (i.e., Eq.~\ref{eq:lat_pooled}) indicate positive serial correlation in the residuals (Durbin–Watson $\simeq 0.71$ for RLS and 1.1 for OLA; note that values $\ll 2$ indicate positive autocorrelation). This is expected given that the data used to estimate $d_{\max}$ were smoothed with a 1-yr running mean (13 Carrington rotations) and sampled in one-Carrington rotation increments to suppress short-timescale noise while preserving solar-cycle variations (Section~\ref{sec:data}); this induces short-range dependences. Accordingly, all quoted standard errors were computed using Newey–West heteroskedasticity- and autocorrelation-consistent (HAC) standard errors \citep[e.g.][]{hamilton1994}.

For both the full and latitude-offset–corrected cases, replacing $\ln(\mathrm{MAI})$ with MAI yields only a slight decrease in $R^2$ (see Table~\ref{tab:dmax_corr}). However, 
the AIC\footnote{AIC (Akaike Information Criterion) is a statistical metric used to compare and select models by balancing goodness-of-fit with simplicity, defined as $\mathrm{AIC}=2k-2\ln(L),$ where $k$ is the number of parameters and $L$ is the maximum likelihood. Lower AIC values indicate better models since the formula penalizes overparameterization.} 
strongly favors the logarithmic model: replacing $\ln(\mathrm{MAI})$ with MAI increases AIC by $\Delta\mathrm{AIC}=+32$ for RLS results, and $+88$ for OLA results. Differences in $\Delta\mathrm{AIC}\gtrsim 10$ are generally considered decisive \citep{burnham2002}, and hence these values provide strong evidence against a linear dependence on MAI. Thus, while a small change in $R^2$ suggests that there is only modest departure from linearity, the change in AIC indicates strong support for the logarithmic rather than linear dependence on MAI.

In summary, both inversion methods show a clear, highly significant negative relationship between $d_{\max}$ and surface magnetic activity. The magnitudes of the inferred slopes differ systematically between the RLS and OLA inversions; this is most likely because OLA results have lower radial resolution than RLS results.

\begin{table*}
\centering
\caption{Comparison of linear regressions of \(d_{\max}\), \(A_{\max}\), and \(\mathrm{FW80}\) against MAI and \(\ln(\mathrm{MAI})\), after subtracting the fitted latitude-specific intercepts \(c_{0,Y}(\theta)\). Columns (left to right) list the inversion method; regressor; slope with Newey–West heteroskedasticity- and autocorrelation-consistent (HAC) standard error for lag of 13 in parentheses; the corresponding \(t\) statistic (\(t_{\mathrm{stat}}=\) slope divided by HAC SE); \(R^2\) from the ordinary least-squares fit; Pearson \(r\); and Spearman \(\rho\). HAC \(p\)-values for the slope are \(\ll 10^{-6}\) for all fits, except for \(\mathrm{FW80}\) using OLA where \(p<5\times 10^{-5}\). The MAI is in units of Gauss, and hence $\ln(\mathrm{MAI})$ is $\ln(\mathrm{MAI})/\mathrm{G}$. Thus for $d_{\max}$, slopes are respectively in units of Mm/$\ln(\mathrm{MAI})$ or Mm/G for the logarithmic and linear fits to the MAI. For $A_{\max}$ (which is a dimensionless quantity), the units are per unit \(\ln(\mathrm{MAI})\) and per unit Gauss.
}
\label{tab:dmax_corr}
\begin{tabular}{lllcccc} 
\hline
Inversion & Regressor & Slope (HAC SE) & $t_{\mathrm{stat}}$ & $R^2$ & Pearson & Spearman\\
Method & & & & & $r$ & $\rho$\\
\hline
\multicolumn{7}{c}{\textbf{$d_{\max}$}}\\
\hline
RLS  & $\ln(\mathrm{MAI})$ & -0.0969 (0.0040) & -24 & 0.82 & -0.72 & -0.71 \\
OLA  & $\ln(\mathrm{MAI})$ & -0.0725 (0.0027) & -27 & 0.79 & -0.69 & -0.69 \\
RLS  & MAI        & -0.0166 (0.0007) & -23 & 0.74 & -0.72 & -0.71 \\
OLA  & MAI        & -0.0120 (0.0005) & -23 & 0.67 & -0.66 & -0.69 \\
\hline
\multicolumn{7}{c}{\textbf{$A_{\max}$}}\\
\hline
RLS  & $\ln(\mathrm{MAI})$ & -0.2195 (0.0072) & -31 & 0.67 & -0.57 & -0.59 \\
OLA  & $\ln(\mathrm{MAI})$ & -0.1169 (0.0057) & -21 & 0.39 & -0.36 & -0.35 \\
RLS  & MAI        & -0.0393 (0.0013) & -31 & 0.59 & -0.59 & -0.59 \\
OLA  & MAI        & -0.0198 (0.0013) & -15 & 0.28 & -0.36 & -0.35 \\
\hline
\multicolumn{7}{c}{\textbf{$\mathrm{FW80}$}} \\
\hline
RLS  & $\ln(\mathrm{MAI})$ & -0.0263 (0.0010) & -26 & 0.53 & -0.46 & -0.46 \\
OLA  & $\ln(\mathrm{MAI})$ & -0.0151 (0.0021) & -7  & 0.08 & -0.14 & -0.13 \\
RLS & MAI         & -0.0042 (0.0002) & -18 & 0.38 & -0.43 & -0.46 \\
OLA &  MAI        & -0.0021 (0.0005) & -5 & 0.03 & -0.11 & -0.13\\
\hline
\end{tabular}
\end{table*}

\paragraph{\bf Behavior of $A_{\max}$}
The absolute value of the amplitude of the maximum shear, $A_{\max}$, increases with MAI (middle panel of Fig.~\ref{fig:rmax_mai_clatns_mai}). 
The middle row of Fig.\ref{fig:rmax_mai_clatns_time} shows $A_{\max}$ as a function of time at two latitudes. The latitude-specific intercepts $c_{0,A_{\max}}(\theta)$ span approximately $-3.6$ to $-2.7$ for OLA and RLS within \(\pm 22.5^\circ\). 

The fitted coefficients are highly significant ($|t|\gtrsim 90$ for the intercepts and $t_{\mathrm{stat}}\simeq 11$ for the common slope $c_{1,A_{\max}}$). Subtracting $c_{0,A_{\max}}(\theta)$ and fitting a slope-only linear model to the residuals leaves the slope essentially unchanged for results obtained by both inversion methods; the corrected fit yields $R^2\simeq 0.7$ for RLS and $0.4$ for OLA (Table~\ref{tab:dmax_corr}).
As expected from the differences in $R^2$, the correlation is weaker for OLA (Pearson/Spearman correlation coefficient of $\approx -0.35$) than for RLS ( coefficients $\approx -0.6$), but remains highly significant in both cases ($|t_{\mathrm{stat}}|\simeq 20$ for OLA and $\simeq 30$ for RLS). Diagnostic tests do not indicate appreciable serial correlation in these fits (Durbin–Watson statistic $\approx 2$).

If we do the regression analysis with MAI rather than $\ln(\mathrm{MAI})$, we find that whether MAI or $\ln(\mathrm{MAI})$ gives a better fit depends on the inversion method. For OLA, the $\ln(\mathrm{MAI})$ model gives higher $R^2$ (0.39 vs. 0.28) and lower AIC ($\Delta\mathrm{AIC}\approx -8$ relative to the value obtained with MAI). For RLS, $R^2$ is also higher for $\ln(\mathrm{MAI})$ (0.67 vs. 0.59), but AIC favors MAI by $\approx 38$. Given this inconsistency, we conclude that a departure from linearity in the $A_{\max}$–MAI relation is not clearly established.

We also find that the inferred slopes differ by nearly a factor of two between OLA and RLS. This discrepancy likely reflects, at least in part, sensitivity to the inversion trade-off (regularization) parameters. Thus, while the sign and qualitative dependence of $A_{\max}$ on surface activity are robust, the values of the slope should be interpreted with caution because of their dependence on the inversion procedure.

\paragraph{\bf{Behavior of \(\mathrm{FW80}\)}}
For $\mathrm{FW80}$ (i.e., the width of the region with strong shear), the RLS results are shown in the bottom panels of Fig.~\ref{fig:rmax_mai_clatns_mai} and~\ref{fig:rmax_mai_clatns_time}; OLA results are very similar. The latitude-dependent intercepts 
$c_{0,{\mathrm{FW80}}}(\theta)$ are \(\sim 0.9\)–1.0~Mm and are highly significant ($|t_{\mathrm{stat}}|\gtrsim 80$ for each intercept in the latitude range within $\pm 22.5^\circ$). The common slope $c_{1,{\mathrm{FW80}}}$ is negative for both inversion methods, indicating that higher MAI is associated with a modest decrease in $\mathrm{FW80}$. 
The slope of the fits to Eq.~\ref{eq:lat_pooled} is statistically significant ($t_{\mathrm{stat}}\simeq 11$ for RLS results and $t_{\mathrm{stat}}\simeq 4$ for OLA results). After subtracting the fitted latitude-dependent intercept, $c_{1,{\mathrm{FW80}}}$, and fitting a slope-only model, we obtain an essentially unchanged slope, but one that is even more statistically significant ($t_{\mathrm{stat}}\simeq 26$ for RLS and $t_{\mathrm{stat}}\simeq 7$ for OLA); as shown in Table~\ref{tab:dmax_corr}. The corresponding correlation coefficients are $r\simeq -0.46$ (RLS) and $-0.14$ (OLA). Consistent with these weak correlations, the explained variance is small for OLA ($R^2\lesssim 0.1$) and moderate for RLS ($R^2\simeq 0.53$).
Durbin–Watson statistics ($\simeq 1.7$ for RLS and $\simeq 2.0$ for OLA) do not indicate strong residual autocorrelation. Overall, RLS shows a statistically significant but moderate negative association between $\mathrm{FW80}$ and activity, whereas OLA shows only a weak relationship despite formal significance.
Replacing $\ln(\mathrm{MAI})$ with MAI yields a poorer fit. For RLS, $R^2$ decreases and AIC increases by 42, decisively favoring the fits to $\ln(\mathrm{MAI})$. For OLA, the change is smaller ($R^2$ decreases from 0.08 to 0.03 and AIC increases by 8). Thus, results of both inversion methods favor a logarithmic dependence of $\mathrm{FW80}$ on MAI, with substantially stronger evidence for RLS.

\begin{figure*}
    \includegraphics[width=0.32\linewidth]{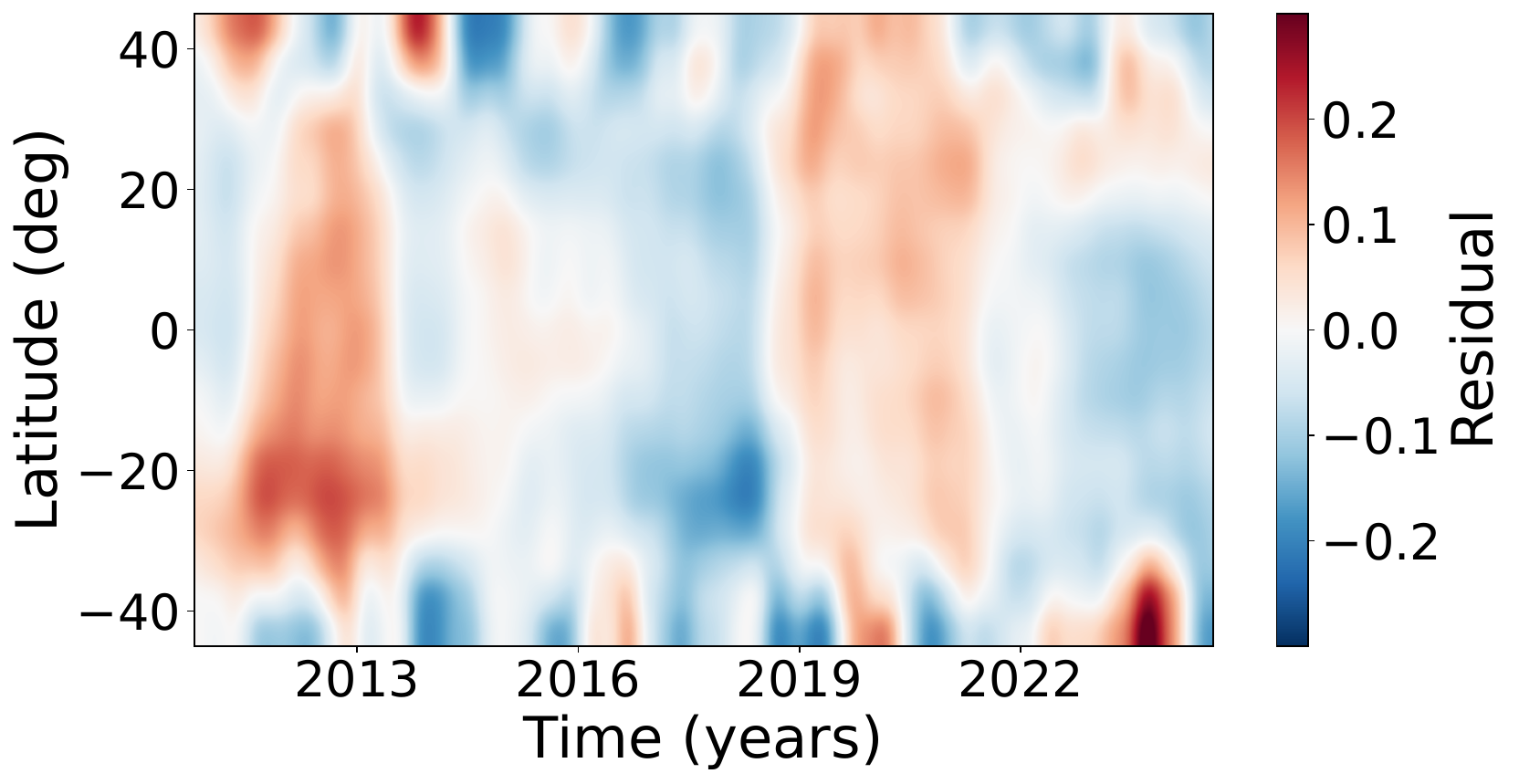}
    \includegraphics[width=0.32\linewidth]{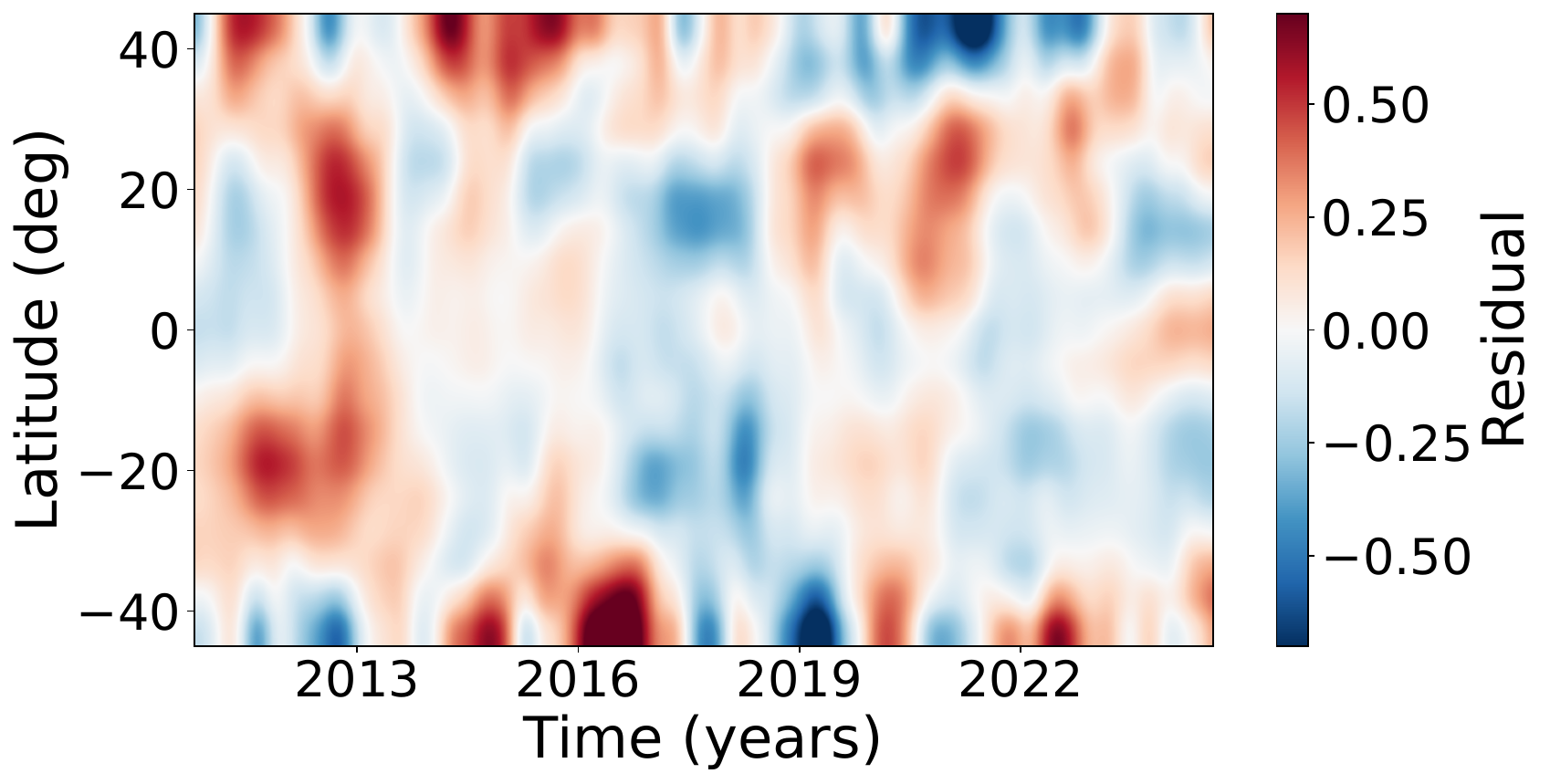}
    \includegraphics[width=0.32\linewidth]{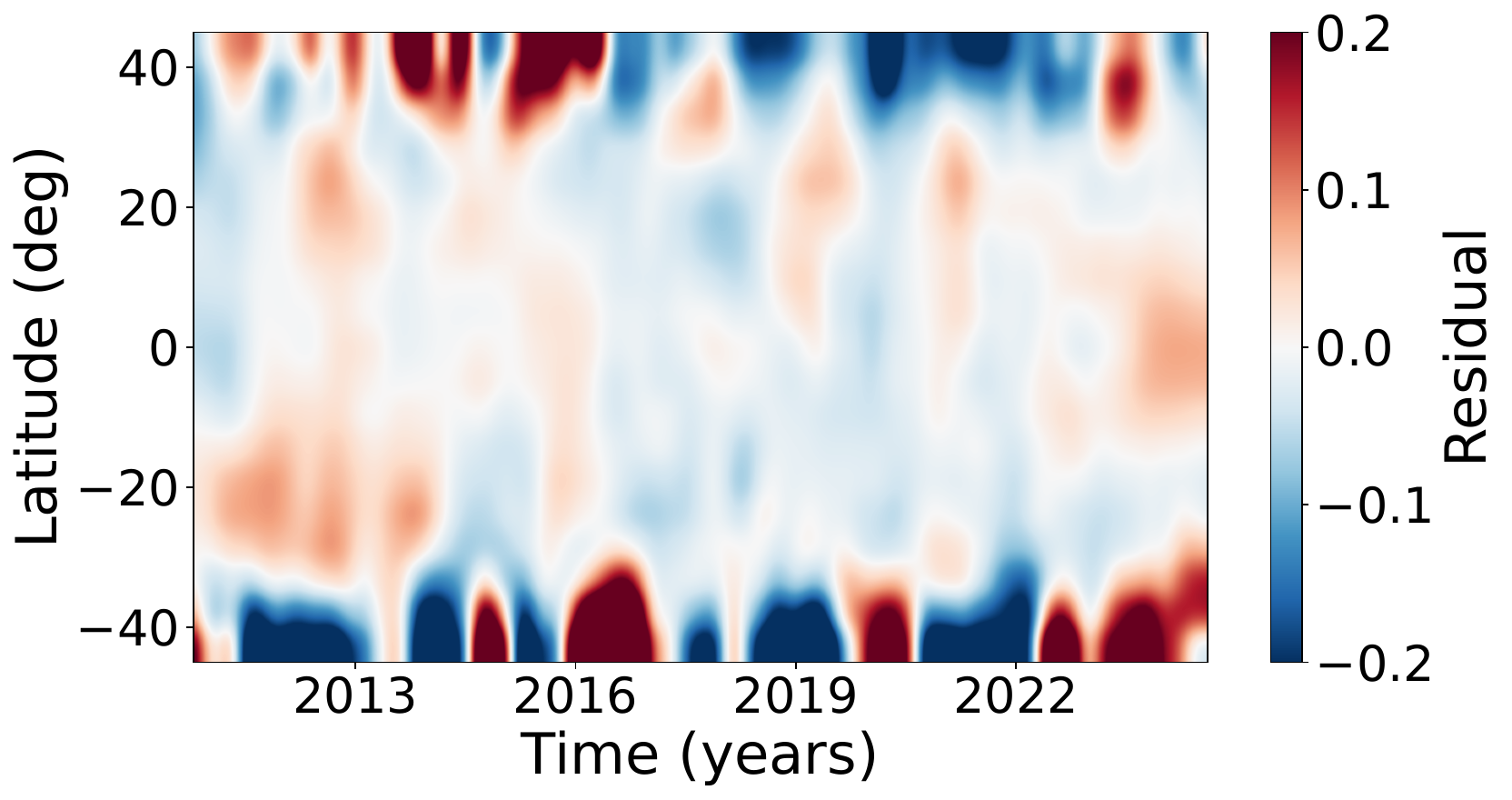}
    \includegraphics[width=0.32\linewidth]{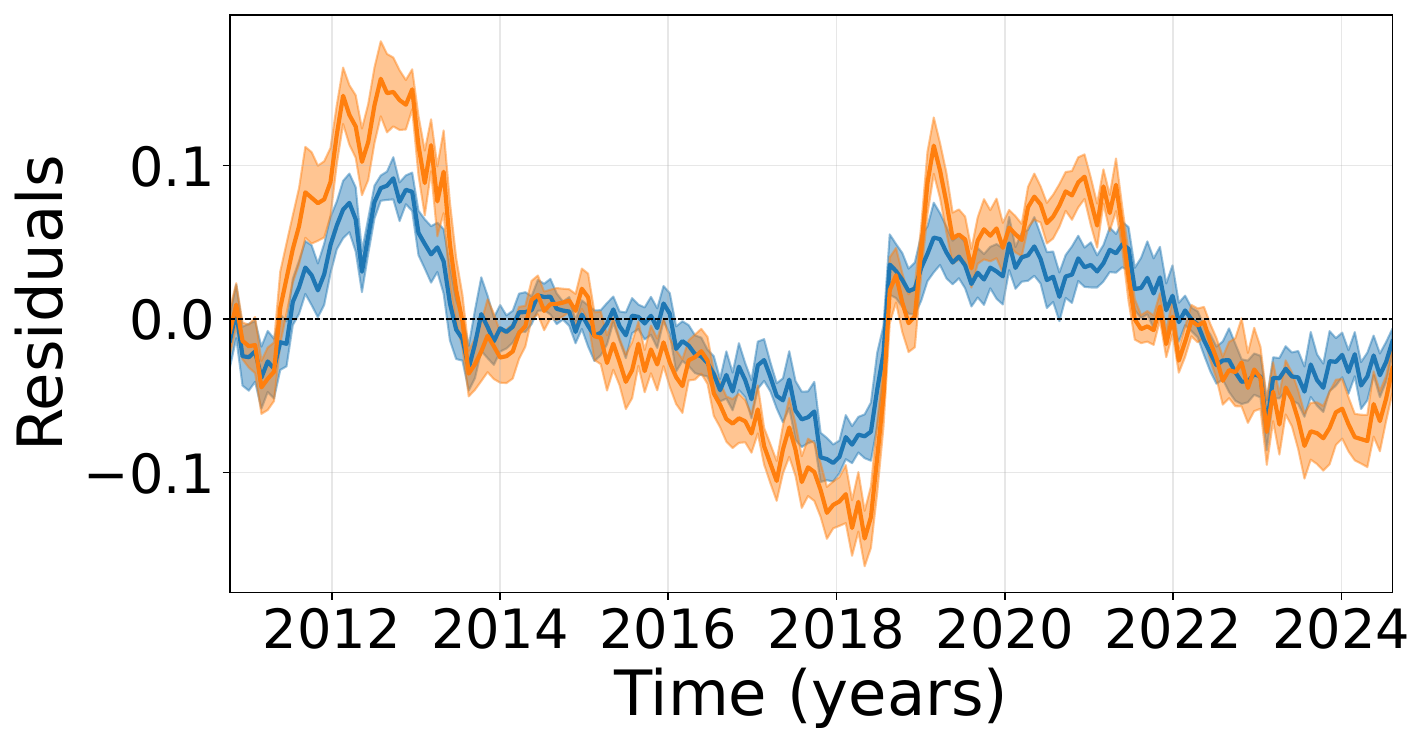}
    \includegraphics[width=0.32\linewidth]{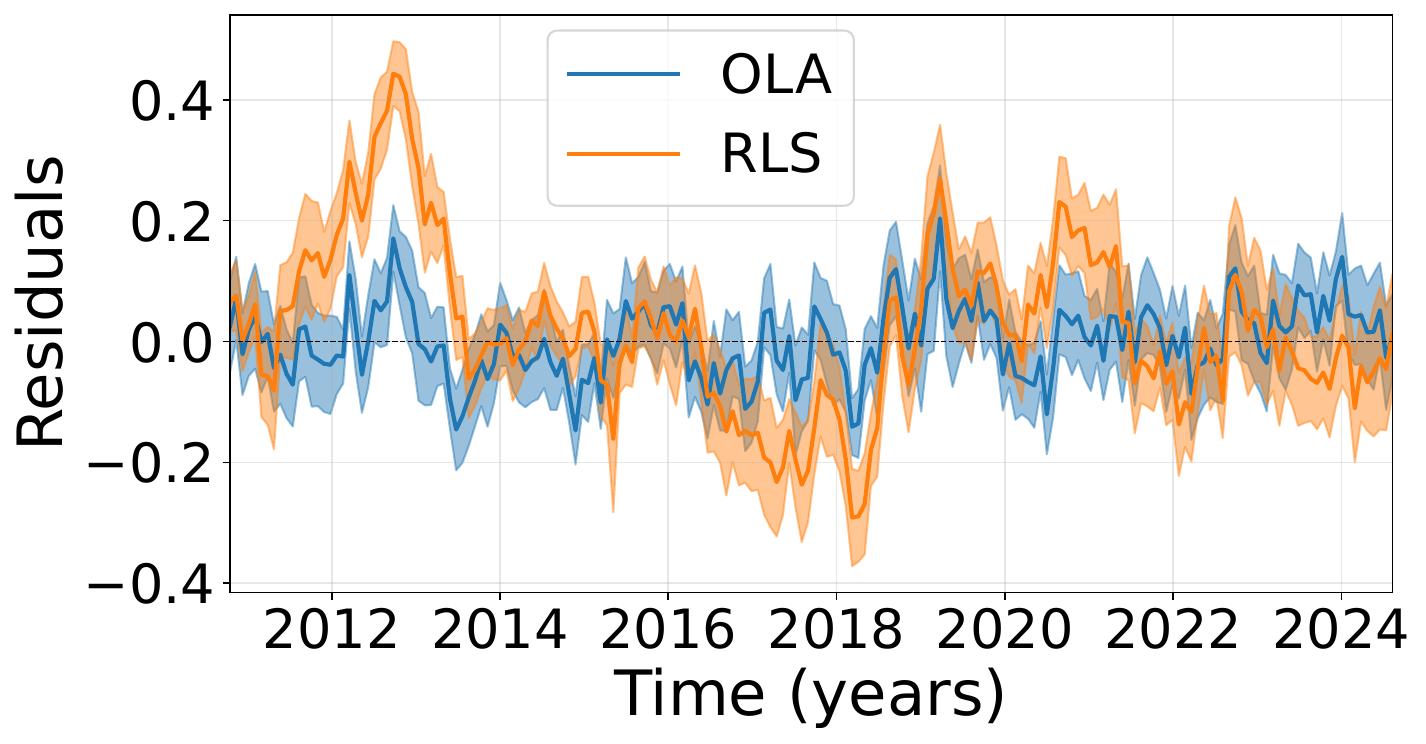}
    \includegraphics[width=0.32\linewidth]{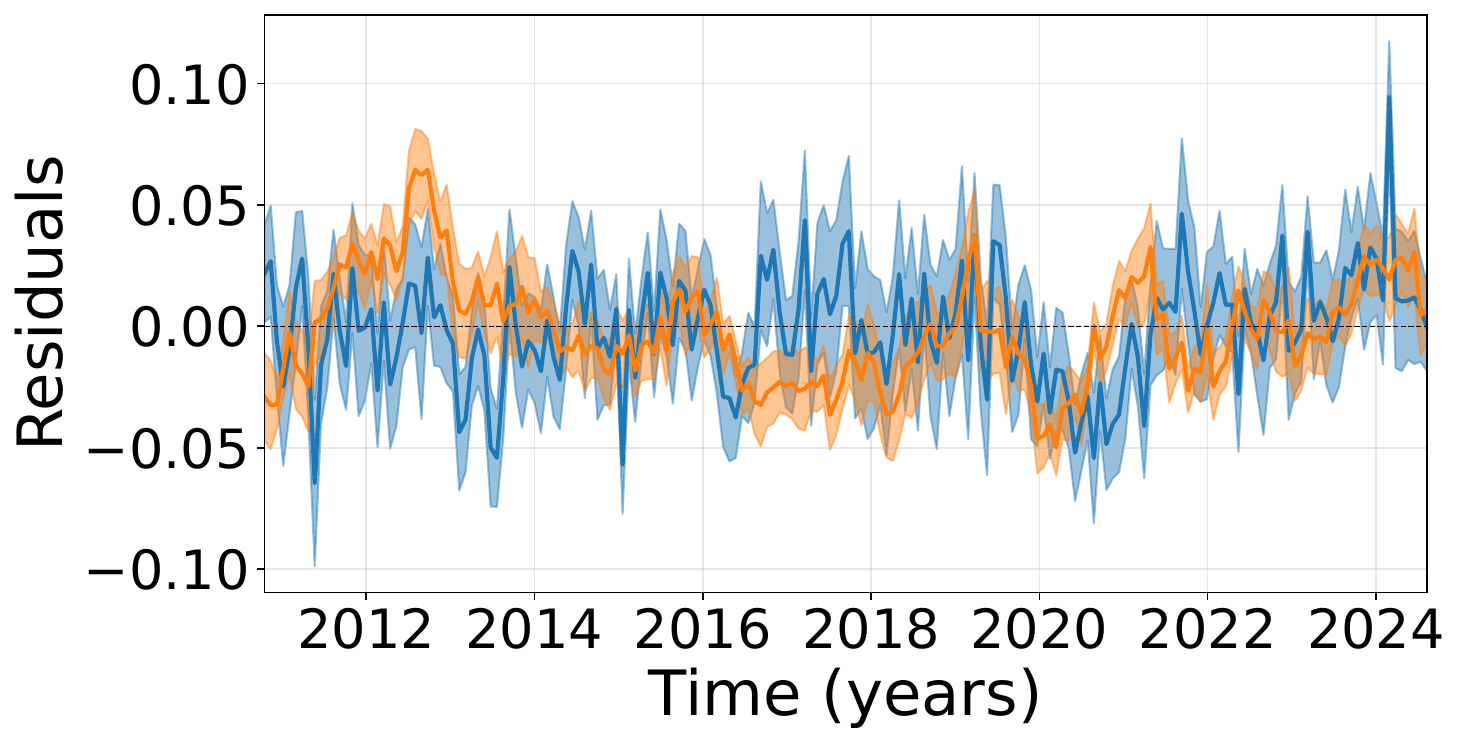}
    \caption{
    Residuals of \(d_{\max}\) (Mm), \(A_{\max}\), and \(\mathrm{FW80}\) (Mm) (left, middle, and right columns, respectively) after removing the fit to Equation~\ref{eq:lat_pooled}. Top: RLS residuals smoothed with a two-dimensional Gaussian filter ($\sigma=7.5^\circ$ in latitude and 2 Carrington rotations in time). Bottom: Time series of the residuals averaged over a latitude range of $\pm 45^\circ$; the shaded region indicates the standard error of the mean. The bottom panels show both RLS and OLA. Outliers exceeding $\pm 2.5\sigma$ at all latitudes were removed prior to taking the latitudinal average, and the gaps were filled by a linear interpolation in time.  
    }
    \label{fig:residuals}
\end{figure*}

\begin{figure}
	\includegraphics[width=\columnwidth]{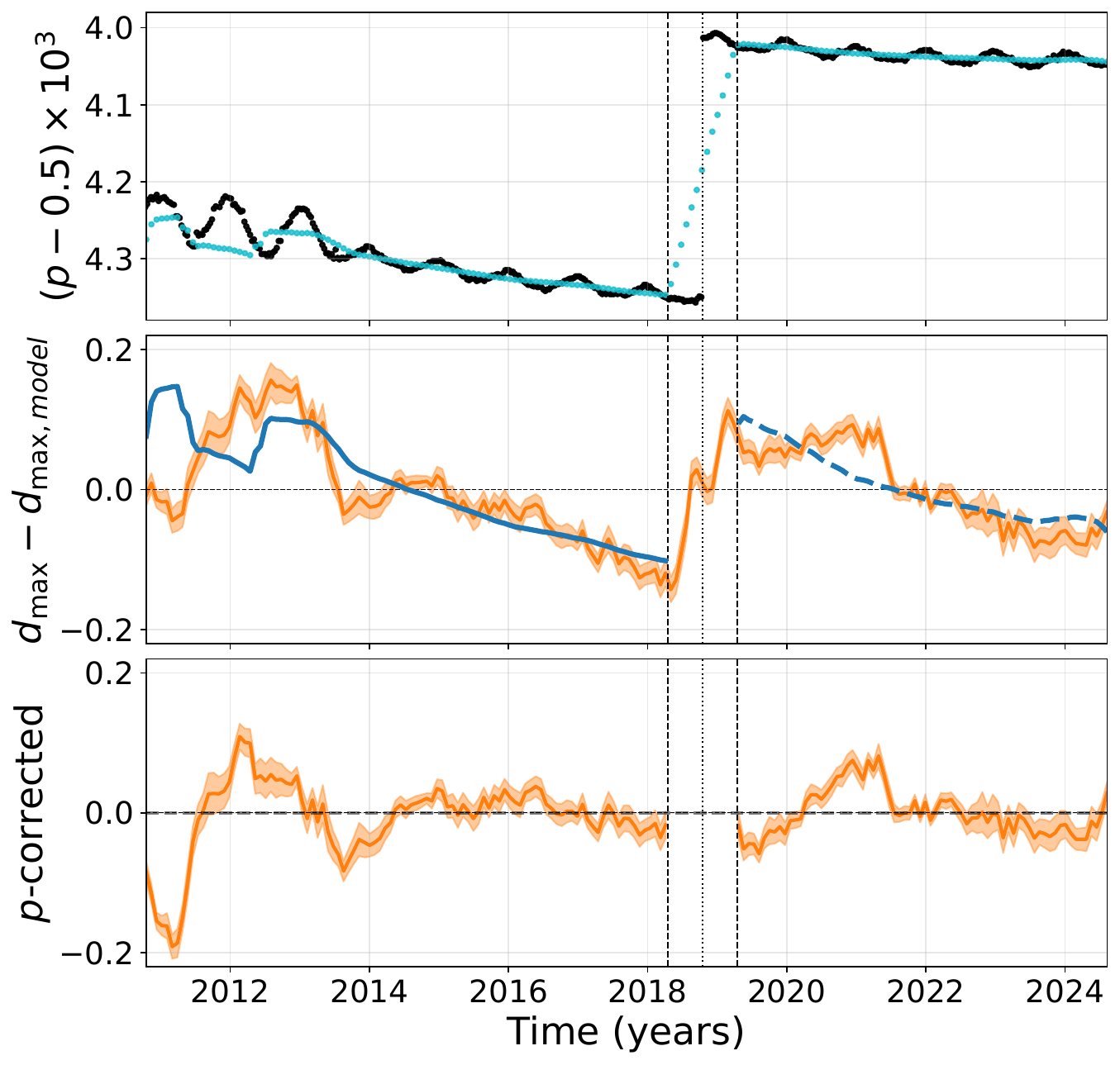}
   \caption{
   Top: HMI plate scale \(p\) plotted as a function of time. The black circles are measurements with a 10-day cadence; the blue curve shows a 1 yr running mean. The nominal plate scale is \(p\simeq 0.5~\mathrm{arcsec~pixel^{-1}}\).
   The vertical dashed lines indicate times 6 months before and after the 2018.7 focus adjustment, which coincides with a step-like change in plate scale.
   Middle: The \(d_{\max}\) residuals (Mm) from the RLS inversion plotted as a function of time.
   The curves show predictions from linear regressions of the residuals against the 1 yr-smoothed plate scale, fit separately before (solid) and after (dashed) the 2018.7 focus change. 
   Bottom: Remnant residuals after removing the plate-scale regression.
   }
    \label{fig:rmax_residuals_scale}
\end{figure}

\paragraph{{\bf Behavior of the residuals}}

Fig.~\ref{fig:residuals} shows the residuals in $d_{\max}$, $A_{\max}$ and \(\mathrm{FW80}\) in the left, middle and right panels respectively. The residuals were obtained by subtracting the fitted model (Eq.~\ref{eq:lat_pooled}) from each parameter, where the combined regression was performed across $|\theta|\le 67.5^\circ$. The upper panels show the residuals as a function of latitude and time, and as can be seen, they are not distributed randomly, at least for $d_{\max}$. To examine the non-randomness, and since the features appear to be more-or-less independent of latitude, we averaged the residuals in the latitude range of $\pm 45^\circ$ and plotted them as a function of time (Fig.~\ref{fig:residuals}, bottom row). Including higher latitudes yields a similar result, but with somewhat higher noise. 

For $A_{\max}$, the residuals differ substantially between the two inversion methods: the RLS residuals show a pattern similar to that in $d_{\max}$, whereas the OLA residuals show no clear periodicity and are consistent with noise. The $\mathrm{FW80}$ residuals are consistent with random noise for both inversions. 

The latitude-averaged $d_{\max}$ residuals suggest a quasi-periodic signal. A Lomb–Scargle periodogram shows a prominent peak near 8~yr (power $\approx 0.6$), well above the 99th-percentile level obtained from time-shuffled realizations ($\approx 0.06$). However, the observational record spans only 13.8~yr, so we regard this $\sim 8$~yr signal as tentative.

Since a sudden sharp change in the residuals is rather surprising, we tried to assess whether potential instrumental effects could cause this. In particular, we examined the evolution of the HMI plate scale and found that the step-like change associated with an October 2018 change in focus control coincides with an abrupt change in the residuals, suggesting that at least part of the apparent long-period signal may be instrumental. Fig.~\ref{fig:rmax_residuals_scale} (top) shows the plate scale over the SDO mission, where the blue curve is a 1~yr running mean, matching the smoothing used in our analysis.
In addition to this step-like feature, the plate scale exhibits a long-term increase over time. Fig.~\ref{fig:rmax_residuals_scale} (middle) shows linear fits of the $d_{\max}$ residuals as a function of the plate scale before (solid) and after (dashed) the focus change.
We find that $d_{\max}$ varies with the HMI plate scale: an increase (decrease) in image scale of a few \(\mathrm{mas\,pixel^{-1}}\) corresponds to a shift in \(d_{\max}\) of a few tenths of a Mm toward the solar surface (solar interior).
While the change in plate scale is accounted for in our analysis, a small change in the systematic errors would result in small shifts in the measured helioseismic degrees, which would in turn slightly affect the radial distribution of inferred velocities. Some small residual variability remains that is not accounted for by plate-scale variations (Fig.~\ref{fig:rmax_residuals_scale} bottom).

The effect of the 2018 plate-scale step on \(A_{\max}\) and \(\mathrm{FW80}\) can be evaluated from the latitude-averaged residuals in the bottom row of Fig.~\ref{fig:residuals}. While the step-like feature is clear in \(d_{\max}\), we find no comparably clear signature in \(\mathrm{FW80}\); \(A_{\max}\) shows at most a weak indication in the RLS results that is not reproduced by OLA, and we therefore regard the evidence as inconclusive. Given that  \(A_{\max}\) is an amplitude and  \(\mathrm{FW80}\) is a width, one expects that any shift in the radial coordinate caused by plate-scale errors would have small, at worst second order effects on these quantities. In any event, this effect does not appear to affect any of the results of the present work.

\paragraph{{\bf Behavior of the latitude-dependent intercepts}} To examine the behavior of the latitude-dependent offsets $c_{0,Y}(\theta)$, we again fit the same model (Eq.~\ref{eq:lat_pooled}) over the full latitude range ($|\theta|\le 67.5^\circ$) using the ordinary least squares method on the complete sample (i.e., 3420 observations). 
As before, the fits were carried out independently for the OLA and RLS inversions. The results are shown in Table~\ref{tab:fullmodel}. 
The model explains most of the change in $d_{\max}$ ($R^2=0.91$ for RLS and 0.88 for OLA). The fraction of variation explained is moderate for $\mathrm{FW80}$ ($R^2=0.51$ for RLS and 0.64 for OLA), 
while $A_{\max}$ shows a strong method dependence ($R^2=0.30$ for RLS and 0.75 for OLA). The Durbin–Watson statistic is close to 2 for these fits. 
The common slopes $c_{1,Y}$ remain negative, consistent with the low-latitude, activity-belt results (Table~\ref{tab:dmax_corr}). 
However, when we extend the latitude range, the best-fit slopes become slightly smaller in magnitude, indicating a modest attenuation of the $\ln(\mathrm{MAI})$ dependence. The difference between the all-latitude and activity-belt slope estimates does not exceed three times the combined HAC standard error.

\begin{figure}
	\includegraphics[width=\columnwidth]{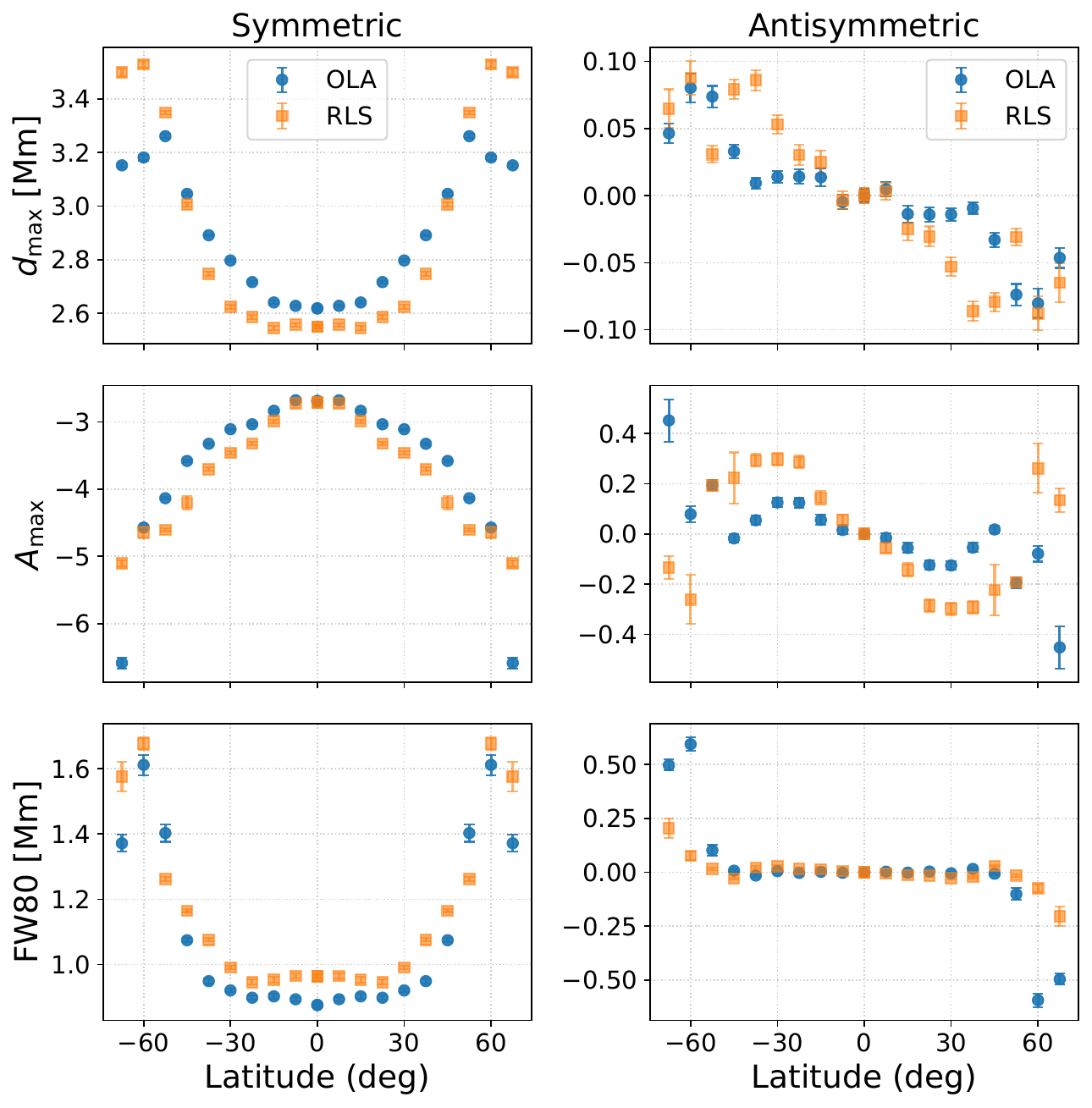}
    \caption{Latitude dependence of the fitted intercepts $c_{0,Y}(\theta)$ for $Y\in\{d_{\max},A_{\max},\mathrm{FW80}\}$. Left: symmetric component; right: antisymmetric component. Vertical error bars show the combined (north–south) uncertainty. 
    See Fig.~\ref{fig:avg_shear} for the corresponding time-averaged shear profiles.
    }
    \label{fig:rmax_amax_c0}
\end{figure}

The intercepts $c_{0,Y}(\theta)$ capture the latitude dependence after accounting for $\ln(\mathrm{MAI})$; their symmetric and antisymmetric components are shown in Fig.~\ref{fig:rmax_amax_c0}. The dominant signal is a systematic increase of $|c_{0,Y}|$ with $|\theta|$ for all three quantities, with a measurable hemispheric offset (the symmetric component dominates, but the antisymmetric component is non-negligible). 

If we average the OLA and RLS intercept trends, the change in the symmetric component between the equator and $52.5^\circ$ is $0.72\pm0.08$~Mm (28$\%\pm$4$\%$) in $d_{\max}$, $1.7\pm0.2$ (62$\%\pm$8$\%$) in $|A_{\max}|$, and $0.4\pm0.1$~Mm (46$\%\pm$15$\%$) in $\mathrm{FW80}$. 
The increase of $|A_{\max}|$ toward higher $|\theta|$ is consistent with the trend observed over $0.991$–$0.996\,R_\odot$ (Fig.~\ref{fig:mean_rotation_shear_sym}, middle panel), corresponding to the radial range of $d_{\max}$, and is opposite to the deeper-layer behavior.
The quoted uncertainties reflect method-dependent systematics (taken as half the OLA–RLS difference); formal propagated fitting errors on the means are much smaller (see Table~\ref{tab:fullmodel}). The latitudinal trends are robust, but their interpretation should account for method-dependent systematics. 

The antisymmetric components of $c_{0,Y}({\theta})$ are smaller than the symmetric component, as seen in Fig.~\ref{fig:rmax_amax_c0}. For $d_{\max}$, the north-south antisymmetric part of the intercept shows an approximately linear dependence on latitude, with slope $\approx -0.001$~Mm deg$^{-1}$. 
For $A_{\max}$, an antisymmetric component is present, but its amplitude and latitude-by-latitude structure differ between OLA and RLS, and hence, the pattern is not robust. For $\mathrm{FW80}$, the antisymmetric signal is appreciable only at the highest latitudes ($|\theta|\ge 52.5^\circ$) and shows substantial discrepancies between inversions; this could be simply because high-latitude flows are difficult to determine.

\begin{table*}[htb]
\centering
\caption{Latitude-specific intercepts $c_{0,Y}(\theta)$ from the pooled fits of Eq.~\ref{eq:lat_pooled} over $|\theta|\le 67.5^\circ$, for $Y\in\{d_{\max},A_{\max},\mathrm{FW80}\}$. Parentheses give Newey–West HAC standard errors (lag $=13$). Values are rounded to three decimal places.}
\label{tab:fullmodel}
\begin{tabular}{r c c c c c c}
\hline
Lat &
\multicolumn{2}{c}{$d_{\max}$} &
\multicolumn{2}{c}{$A_{\max}$} &
\multicolumn{2}{c}{$\mathrm{FW80}$} \\
(deg) &
\multicolumn{2}{c}{(Mm)} &
\multicolumn{2}{c}{} &
\multicolumn{2}{c}{(Mm)} \\
\hline
 & RLS & OLA & RLS & OLA & RLS & OLA \\
\hline
  0.0 & 2.550 (0.008) & 2.618 (0.007) & -2.714 (0.022) & -2.690 (0.020) & 0.964 (0.007) & 0.876 (0.009) \\
  7.5 & 2.561 (0.010) & 2.633 (0.008) & -2.777 (0.031) & -2.694 (0.027) & 0.959 (0.011) & 0.896 (0.009) \\
 -7.5 & 2.554 (0.008) & 2.623 (0.007) & -2.666 (0.026) & -2.663 (0.020) & 0.970 (0.010) & 0.891 (0.009) \\
 15.0 & 2.520 (0.012) & 2.627 (0.010) & -3.134 (0.039) & -2.891 (0.030) & 0.939 (0.013) & 0.900 (0.010) \\
 -15.0 & 2.570 (0.012) & 2.654 (0.009) & -2.848 (0.035) & -2.779 (0.028) & 0.967 (0.012) & 0.905 (0.010) \\
 22.5 & 2.556 (0.009) & 2.702 (0.007) & -3.607 (0.038) & -3.160 (0.028) & 0.931 (0.011) & 0.902 (0.010) \\
 -22.5 & 2.617 (0.012) & 2.731 (0.009) & -3.036 (0.033) & -2.912 (0.025) & 0.961 (0.012) & 0.895 (0.009) \\
 30.0 & 2.572 (0.009) & 2.783 (0.007) & -3.757 (0.032) & -3.235 (0.025) & 0.963 (0.007) & 0.916 (0.008) \\
 -30.0 & 2.678 (0.010) & 2.811 (0.006) & -3.162 (0.038) & -2.985 (0.024) & 1.019 (0.009) & 0.926 (0.008) \\
 37.5 & 2.662 (0.009) & 2.882 (0.006) & -3.996 (0.044) & -3.380 (0.030) & 1.054 (0.007) & 0.964 (0.009) \\
 -37.5 & 2.834 (0.012) & 2.901 (0.006) & -3.412 (0.024) & -3.271 (0.022) & 1.098 (0.008) & 0.934 (0.011) \\
 45.0 & 2.927 (0.012) & 3.013 (0.008) & -4.423 (0.202) & -3.563 (0.024) & 1.195 (0.007) & 1.068 (0.010) \\
 -45.0 & 3.086 (0.008) & 3.079 (0.006) & -3.978 (0.016) & -3.598 (0.020) & 1.135 (0.005) & 1.082 (0.008) \\
 52.5 & 3.319 (0.009) & 3.187 (0.012) & -4.796 (0.027) & -4.329 (0.030) & 1.247 (0.007) & 1.301 (0.019) \\
 -52.5 & 3.380 (0.009) & 3.335 (0.011) & -4.412 (0.031) & -3.942 (0.029) & 1.277 (0.009) & 1.504 (0.050) \\
 60.0 & 3.442 (0.016) & 3.101 (0.011) & -4.376 (0.042) & -4.650 (0.050) & 1.602 (0.028) & 1.018 (0.043) \\
 -60.0 & 3.617 (0.019) & 3.262 (0.018) & -4.897 (0.190) & -4.493 (0.039) & 1.753 (0.024) & 2.206 (0.046) \\
 67.5 & 3.434 (0.018) & 3.106 (0.012) & -4.967 (0.068) & -7.037 (0.150) & 1.372 (0.070) & 0.875 (0.042) \\
 -67.5 & 3.564 (0.022) & 3.199 (0.009) & -5.234 (0.062) & -6.135 (0.079) & 1.780 (0.056) & 1.868 (0.029) \\
\hline
\end{tabular}
\medskip
\end{table*}

To assess whether the changes shown in Fig.~\ref{fig:rmax_amax_c0} are real or an artifact of the finite resolution of the inversions, we looked at how the averaging kernels (i.e., the resolution kernels) affect $d_{\max}$ and $A_{\max}$.
The averaging kernels vary only weakly in time (across Carrington rotations), but they vary slightly with latitude despite being broadly similar in shape.

\begin{figure}
	\includegraphics[width=\linewidth]{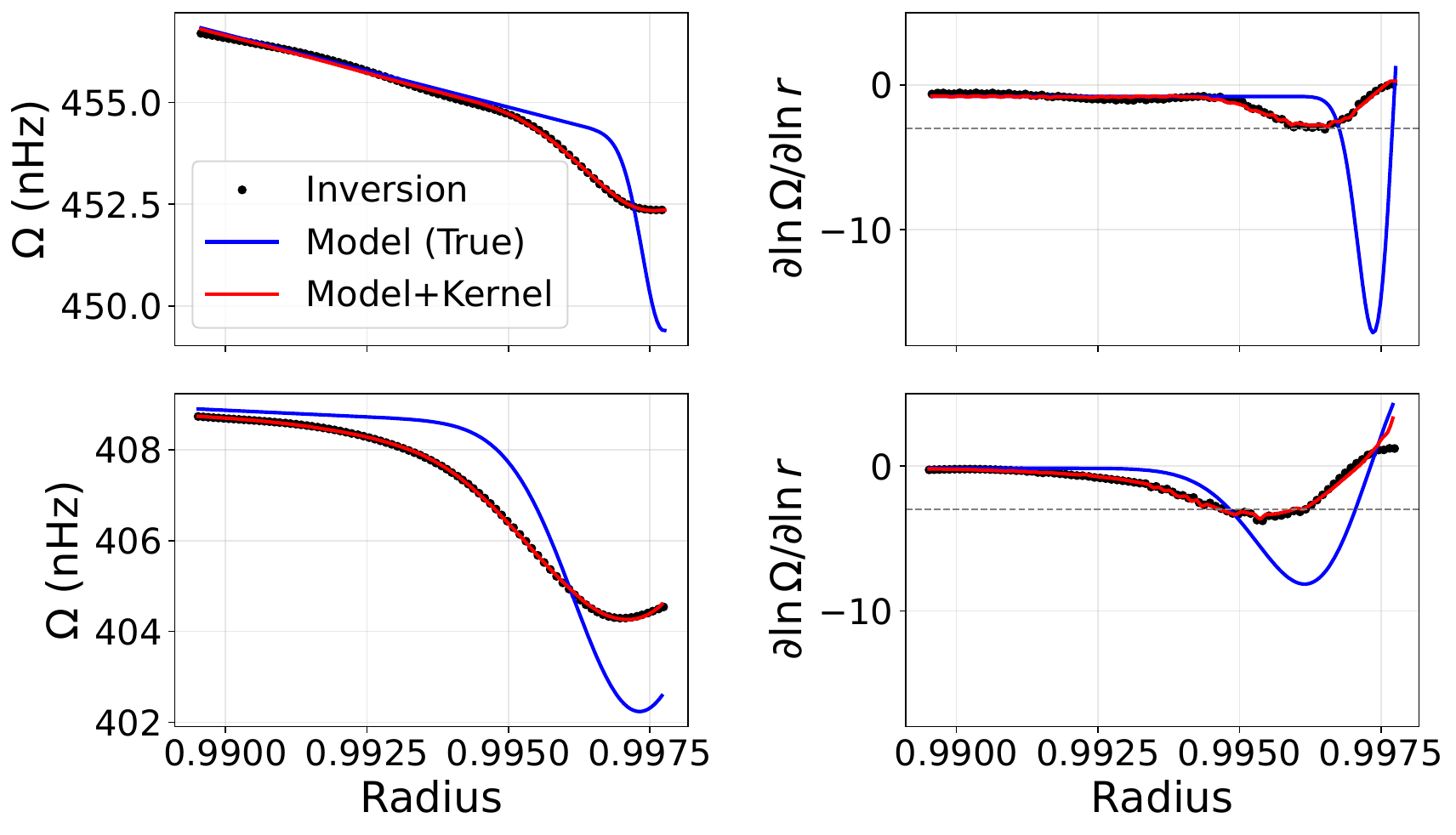}
    \caption{Left: Simulated rotation rate \(\Omega\) (blue) and the finite-resolution result after convolution with the averaging kernel (red). Right: The corresponding shear (blue) and the kernel-convolved shear (red). The top panels are for an equatorial tile, and the bottom panels are for a tile at \(45^\circ\)N. The finite resolution smooths \(\Omega\), reduces the prominence of the shear dip (affecting \(A_{\max}\)), and shifts the inferred dip to greater depth.}
        \label{fig:avk}
\end{figure}

To study the effect of the averaging kernels on our inferences, we adopted the same simple parametric model for the zonal-flow perturbation used by \citet{rabello2024} and propagated the model through the OLA averaging kernels prior to comparison with the inversions. The model consists of a Gaussian perturbation with negative amplitude superposed on a linear background (five free parameters: Gaussian amplitude, width, and center, plus background slope and offset). We convolved the model with the OLA averaging kernels and fit the resulting kernel-weighted profile to the OLA-inverted $\Omega(r)$ using a Levenberg–Marquardt least-squares routine. Fits were performed for three representative Carrington rotations: two during solar minimum (CR 2218 and 2231) and one during solar maximum (CR 2277). We show a sample of the results in Fig.~\ref{fig:avk}.

As noted by \citet{rabello2024}, the broad averaging kernels map a sharp near-surface decrease in rotation rate to appreciably greater depths; consequently, the intrinsic (unconvolved) parametric model places the minimum closer to the surface than the kernel-weighted fit.
At the equator, Table~\ref{tab:fullmodel} gives $d_{\max}=2.550\pm0.008$~Mm, whereas the intrinsic model yields a minimum $\simeq 0.7$~Mm closer to the surface.
Importantly, the qualitative latitude dependence of $d_{\max}$ is robust: both the kernel-weighted fits and the intrinsic model show a shallowing of the perturbation with increasing latitude. The intrinsic model predicts a larger equator-to-$45^\circ$ change ($\simeq 0.8$~Mm; $\simeq 40\%$) than inferred from Fig.~\ref{fig:rmax_amax_c0} ($\simeq 0.5$~Mm; $\simeq 20\%$). In contrast, the strong increase in $|A_{\max}|$ at high latitudes seen in Fig.~\ref{fig:rmax_amax_c0} is not reproduced by this simple model.

\section{Summary and Conclusions}\label{sec:conclusions}

We have conducted a detailed study of radial shear in the near-surface shear layer of the Sun. Our use of ring-diagram technique allowed us to determine the rotation rate between $0.998\,R_\odot$ and $0.975\,R_\odot$. Thus we can study layers shallower than what is accessible to global-mode analysis. In the deeper layers, the shear ($\partial\ln\Omega/\partial\ln r$) is approximately $-1$, and agrees with global-mode results in those layers \citep{barekat2014, antia2022}.
Unlike global-mode analyses, we are able to examine both the north-south symmetric component of the shear, as well as the north-south antisymmetric component. We find that, although the antisymmetric component has a substantially smaller amplitude than the symmetric component (typically \(\lesssim 25\%\) of the symmetric amplitude), it is not negligible compared to the measurement uncertainties 
(reaching up to \(\sim 9\sigma\)).

As in \citet{rabello2024}, we find that the radial shear ($\partial\ln\Omega/\partial\ln r$) in the NSSL has three distinct regions, with the middle region showing increased shear.
The latitude dependence varies strongly with depth, likely contributing to discrepancies among previous results sampling different radial ranges: the shear can become positive or more negative from the equator to high latitudes depending on depth. At \(r\simeq 0.991\)–\(0.996\,R_\odot\) (\(\sim 3\)--6~Mm), the usual latitudinal trend reverses, with the shear becoming more negative at high latitudes than at the equator. The antisymmetric component also grows with \(|\theta|\) and peaks close to the surface (\(r\gtrsim 0.996\,R_\odot\)); comparisons in \citet{rabello2024} suggest that this signal is not dominated by center-to-limb systematics.

In layers accessible to global-mode analysis, our inferred time variation is consistent with global-mode results, but the largest variations occur at shallower depths (\(r\gtrsim 0.99\,R_\odot\)), where the residuals are substantially larger (approximately five times) than at greater depths. Variations in the layer with the strongest shear (layer M) tend to have the opposite sign from those in the shallowest region (layer S).

Finally, the relationship between magnetic activity and shear depends on depth. Near \(\sim 0.99\,R_\odot\), sunspots preferentially occur where the shear is below average \citep[e.g.,][]{barekat2016, antia2022, komm2022}, whereas at shallower depths the residual shear shows the opposite behavior, with enhanced residuals in regions of strong magnetic activity.
This emphasizes that as the shear changes across the near-surface layers, its magnetic association also changes and should be interpreted with explicit attention to the sampled depth range.

We characterize the strong-shear layer (layer M; Fig.~\ref{fig:rmax_amax_definition})
by the location where the shear is strongest, \(d_{\max}\), its amplitude $A_{\max}$, and its width, \(\mathrm{FW80}\), as functions of time for each latitude band. We find that all three quantities are strongly correlated with the magnetic activity index. Increasing magnetic activity is associated with a shallower, stronger, and modestly narrower strong-shear layer; over a solar cycle ($\Delta \mathrm{MAI}\sim 20$ G) we infer changes of order $\sim 0.3$~Mm in $d_{\max}$, $\sim 0.4$–0.8 in $A_{\max}$, and $\sim 0.05$~Mm in $\mathrm{FW80}$.
We fit a simple linear function in $\ln({\mathrm{MAI}})$ to all three quantities and find that for each quantity the slope is the same for all latitudes, but there is a latitude dependence in the intercept.
Toward higher latitudes (\(\gtrsim 52.5^\circ\)), the strong-shear layer shifts to greater depths, becoming stronger and broader (by order of tens of percent) relative to the equator.
We interpret the correlation between $d_{\max}$, $A_{\max}$, and \(\mathrm{FW80}\) with the magnetic index as an indication that the flows and magnetic fields are interconnected in these layers. While this work cannot establish causality, the strength of the correlations suggests physical coupling between magnetic activity and the strong-shear layer.

\citet{Kitchatinov2016} argues that the near‑surface rotational shear should increase with toroidal‑field strength because the magnetic field more strongly quenches turbulent viscosity than it suppresses the $\Lambda$‑effect; although the shear is only weakly sensitive to magnetic fields in general, it can serve as a probe for sufficiently strong fields of order 1~kG \citep{Kitchatinov2016}. 
\citet{baldner2009} inferred two near‑surface toroidal‑field concentrations centered at
$r=0.999\,R_{\odot}$ and $0.996\,R_{\odot}$, with peak amplitudes $380\pm30\,$G and $1.4\pm0.2\,$kG, respectively. The deeper, kG‑scale peak lies at the radius where we observe the strongest shear in layer M, and the coincidence of field strength and shear — together with our finding that the shear increases with sunspot number ($A_{\max}$; see Fig. 6, middle panel) — is consistent with Kitchatinov’s prediction that sufficiently strong toroidal fields can enhance the near‑surface rotational shear (toroidal fields are strongest near activity maximum). We find that the location of the strong shear moves toward the surface during activity maximum (Fig. 6, top panel), which suggests that the strong toroidal field likewise shifts closer to the solar surface at activity maximum.

Two caveats should be kept in mind when interpreting any observational results. First, finite resolution can bias the inferred properties. As noted by \citet{rabello2024}, the broad averaging kernels can shift the apparent location of the strong-shear feature to greater depths; for instance, at the equator, the true location is $\sim 0.7$~Mm closer to the surface.
Although the averaging kernels are largely similar across latitude and nearly uniform over time,
we find that accounting for their effect suggests
that the equator-to-high-latitude increase in \(d_{\max}\) is, if anything, underestimated in the inversions. We also find that
the inferred high-latitude increase in \(|A_{\max}|\) is more sensitive to finite resolution and should be interpreted with caution.
Second, instrumental systematics (e.g., changes in HMI plate scale) can imprint signatures in the time-dependent results and may contribute to some long-period variability.
We find that changes in the HMI plate scale are associated with shifts in \(d_{\max}\) at the level of a few tenths of a Mm for plate-scale variations of a few \(\mathrm{mas\,pixel^{-1}}\).
This underscores the importance of instrument calibration and continuous monitoring; at the same time, the ability to detect such small variations is a testament to the quality of the HMI data and its calibration.

\begin{acknowledgments}
This research was supported by NASA grant 80NSSC25K7669.
This research was also supported in part by NASA Contract NAS5-02139 to Stanford University and by the COFFIES DSC Cooperative Agreement 80NSSC22M0162.
The work uses data from the Helioseismic and Magnetic Imager. HMI data are courtesy of NASA/SDO and the HMI science team. The data used in this article are publicly available from the Joint Science Operations Center at jsoc.stanford.edu.
\end{acknowledgments}

\facilities{HMI(SDO), JSOC(Stanford and Lockheed)}

\bibliography{main}{}
\bibliographystyle{aasjournalv7}

\end{document}